\documentclass[
  twocolumn,
  prb,
  amsmath,
  amssymb,
  superscriptaddress,
  floatfix
]{revtex4}

\usepackage{bm}
\usepackage{graphicx}
\usepackage{dsfont}
\usepackage{float}
\usepackage{cancel}
\usepackage{braket}
\renewcommand{\vec}{\mathbf}

\renewcommand{\Im}{\mathop{\mathrm{Im}}}
\renewcommand{\Re}{\mathop{\mathrm{Re}}}
\renewcommand{\log}{{\,\mathrm{ln}}}                      
\newcommand{\pd}[2]{\frac{\partial #1}{\partial #2}}

\newcommand{\pint}[3]{\int\limits_{#2}^{#3}\! \frac{\mathrm{d}{#1}}{2 \pi}}
\newcommand{\pintd}[4]{\int_{#3}^{#4}\!\! \frac{\mathrm{d}^{#2}#1}{(2\pi)^{#2}}}
\newcommand{\nint}[3]{\int\limits_{#2}^{#3}\! \mathrm{d}{#1}}

\newcommand{\fkt}[2]{#1\!\left(#2\right)}
\newcommand{\fktb}[2]{#1\!\left[#2\right]}
\newcommand{\fktc}[2]{#1\!\left\{#2\right\}}
\newcommand{\ABS}[1]{\left|#1\right|}

\newlength{\graphwid}
\def\showgraph#1#2{
\settowidth{\graphwid}{\includegraphics[#1,clip=true]{#2}}
\parbox[c]{\graphwid}{\includegraphics[#1,clip=true]{#2}}}

\begin{document}

\title{Coulomb interaction in graphene:  Relaxation rates and transport }

\author{M.~Sch\"utt}
\affiliation{
 Institut f\"ur Nanotechnologie, Karlsruhe Institute of Technology,
 76021 Karlsruhe, Germany
}

\author{P.~M.~Ostrovsky}
\affiliation{
 Institut f\"ur Nanotechnologie, Karlsruhe Institute of Technology,
 76021 Karlsruhe, Germany
}
\affiliation{
 L.~D.~Landau Institute for Theoretical Physics RAS,
 119334 Moscow, Russia
}

\author{I.~V.~Gornyi}
\affiliation{
 Institut f\"ur Nanotechnologie, Karlsruhe Institute of Technology,
 76021 Karlsruhe, Germany
}
\affiliation{
 A.F.~Ioffe Physico-Technical Institute,
 194021 St.~Petersburg, Russia.
}

\author{A.~D.~Mirlin}
\affiliation{
 Institut f\"ur Nanotechnologie, Karlsruhe Institute of Technology,
 76021 Karlsruhe, Germany
}
\affiliation{
 Institut f\"ur Theorie der kondensierten Materie, Karlsruhe Institute of Technology,
 76128 Karlsruhe, Germany
}
\affiliation{
 Petersburg Nuclear Physics Institute,
 188350 St.~Petersburg, Russia.
}

\date{\today}

\begin{abstract}
We analyze the inelastic electron-electron scattering in undoped graphene
within the Keldysh diagrammatic approach.  We demonstrate that finite
   temperature strongly affects the screening properties of graphene,
   which, in turn, influences the inelastic scattering rates as
   compared to the zero-temperature case. Focussing on the clean regime, we calculate the
   quantum scattering rate which
is relevant for dephasing of interference processes. We identify an
hierarchy of regimes
arising due to the interplay of a plasmon enhancement of the scattering
and finite-temperature
screening of the interaction.  
We further address the energy relaxation and transport
   scattering rates in graphene. We find a non-monotonic energy dependence of the
inelastic relaxation rates in clean graphene which is attributed to the resonant excitation of plasmons.
Finally, we discuss the temperature dependence of the conductivity at the Dirac point
in the presence of both interaction and disorder.
Our results complement the kinetic-equation and hydrodynamic
approaches for the collision-limited conductivity of clean graphene and can be generalized
to the treatment of physics of inelastic processes
in strongly non-equilibrium setups.
\end{abstract}

\maketitle

\section{Introduction}
\label{SecInt}
Graphene \cite{geim07,graphene-review} is a two-dimensional (2D) material
with a quasi-relativistic dispersion law that has attracted an
outstanding attention of leading experimental as well as
theoretical groups all over the world.
In 2004, researchers at Manchester
University first succeeded  in experimental isolation of
a monoatomic graphite layer
---graphene---on an insulating substrate
\cite{novoselov04,novoselov05a}. This technological breakthrough
was immediately followed by transport
measurements\cite{novoselov05b,kim05} which have shown
  remarkable properties related to Dirac nature of the charge
  carriers.
In particular, a short and wide sample of
clean graphene exhibits a pseudo-diffusive charge transport
\cite{Katsnelson}, with the counting statistics equivalent to that of a
diffusive wire \cite{Tworzydlo06Beenakker08rev, Ludwig}. This equivalence has
been confirmed in recent measurements of conductance and noise in ballistic
graphene flakes \cite{Miao07, Danneau08}. In contrast to conventional metals,
ballistic graphene near the Dirac point conducts better when potential
impurities are added \cite{Titov07, Bardarson07, Schuessler09,titov09}.

Quantum
interference in disordered graphene is also highly peculiar due to Dirac nature
of carriers. In particular, at the Dirac point, the
minimal conductivity\cite{novoselov05b,kim05} $\sim e^2/h$
is ``protected'' from quantum localization in the absence of
intervalley scattering  \cite{OurPapers} or in the case of a
chiral-symmetric disorder \cite{ostrovsky06,ostrovsky10}.
Away from Dirac point, the
concentration dependence of graphene conductivity in diffusive samples
depends strongly on the nature of scatterers \cite{ostrovsky06}. The
experimentally observed (approximately linear) dependence in most of the
samples may be explained by strong impurities creating
resonances near the Dirac point (``midgap states'')
\cite{ostrovsky06,Stauber07}, yielding  $\sigma \propto n \ln^2
n$. Alternative candidates are Coulomb impurities and/or ripples,
leading to\cite{Ando06,Nomura06,ostrovsky06,Morpurgo06}
$\sigma \propto n$. The dominant type (or types) of
disorder and the corresponding disorder strength depend, of course,
on technology of the sample preparation.

How important is the electron-electron interaction in graphene?
Recent works demonstrated
manifestation of the interaction in dephasing rate
providing the cut-off to quantum interference phenomena
\cite{gorbachev08,Tikhonenko09,mccann06}, as well as in renormalization of conductivity
\cite{cheianov06,kozikov10}.
Recent experiments also showed that the interaction plays a particularly
prominent role in suspended graphene samples.\cite{andrei08,bolotin08}
In such
samples the splitting of integer quantum Hall transition (attributed
to interaction-induced spin/valley symmetry breaking) is observed at
magnetic fields as low as 2T (i.e. an order of magnitude less than in
graphene structures on a substrate). Furthermore, one observes also
fractional quantum Hall plateaus, which indicates the importance of
electron correlations.\cite{EAndrei09,bolotin09,andrei09}
Continuous advances in fabrication of high-quality graphene samples are
expected to lead to further enhancement of the role of interaction in
graphene.

On the theoretical side,
interactions may have a dramatic impact on quantum electronic
transport, especially in systems of reduced dimensionality
\cite{AA}. Very generally, the interaction-induced phenomena may be
subdivided in two big classes, related to effects of renormalization
and inelastic scattering, respectively.
In the case of graphene, the interaction physics becomes an even more
complex problem, in view of the ``relativistic'' dispersion of carriers.
Interaction phenomena are particularly strong near the Dirac point,
where the density of states vanishes (for a clean system), the
screening by intrinsic carriers becomes very inefficient, and
the Drude conductivity (in the presence of disorder)
is of the order of the conductance quantum.

The dimensionless bare coupling constant $\alpha_g=e^2/v_F$ describing the Coulomb repulsion
in graphene samples on a SiO$_2$ insulating substrate
is estimated to be $0.6-0.8$
and can be yet larger, $\alpha_g\simeq 2.2,$ in suspended graphene sheets.
Transport experiments on the interaction correction to conductivity in
graphene on a substrate~\cite{kozikov10} have found 
much smaller values of the interaction constant (or, equivalently,
the interaction parameter $r_s$)  which can be attributed
to the renormalization effects which we briefly overview below.
Furthermore, the value of the effective fine structure constant of freestanding graphene
as inferred from the results of recent experiments on X-ray scattering 
in graphite~\cite{Reed10} was also found to be much smaller than the bare value of $\alpha$.

The interaction effects in the clean graphene have been considered in
Refs.~\onlinecite{Gonzalez94,Ye98,Ye99,Stauber05,Son07,Sheehy,Foster08}
within the weak-coupling renormalization-group (RG) scheme
justified for a large number of flavors of Dirac fermions.
The main result of this consideration~\cite{Gonzalez94,Son07} is the
renormalization of the Fermi velocity
and hence
of the interaction parameter.
This perturbative renormalization group for graphene thus shows that in the
clean case the Coulomb interaction
is marginally irrelevant. In the disordered case, a unified ballistic
RG emerges describing renormalization
of disorder couplings
and of the interaction\cite{Ye98,Ye99,Stauber05,Foster06,Sheehy,Foster08}.
In Ref.~\onlinecite{Foster08} the corresponding one-loop RG
equations are derived for time-reversal-invariant disorder and in the limit of
large number of valleys.

At sufficiently strong $\alpha_g$, the Coulomb interaction in a clean graphene
has been argued to give rise to various
instabilities~\cite{Khveshchenko,Gusynin02,Herbut,
Aleiner07,Hands08,Drut09,liu09},
in particular, to opening the gap in the Dirac spectrum of the quasiparticles
(spontaneous mass generation). Physically, this instability leads to a phase
transition from the semimetallic
state to an insulator state, known as excitonic semimetal-insulator
transition~\cite{Khveshchenko}.
Recently, this type of instability has been studied by effective
mean-field-type approach \cite{Khveshchenko,Gusynin02},
renormalization group method \cite{Son07}, and lattice simulations
\cite{Hands08, Drut09}.
The mean-field consideration of the excitonic-type instability in graphene
predicts a certain value of the coupling constant $\alpha_g\simeq 2$ (close to
that in a suspended graphene)
at which the instability occurs. The predicted semimetal-insulator transition,
however, has not
yet been observed in experiments in zero magnetic field. One of the possible
reasons for that
might be the presence of disorder \cite{liu09}.
In strong magnetic field---in the quantum Hall effect regime---
the repulsive interaction between electrons may result in the
Stoner instability \cite{Haldane07, NomuraMcDonald06, abanin07} giving rise to
spontaneous breaking of spin and/or valley symmetry. Experiments do
show splitting of quantum Hall plateaus in strong magnetic fields
\cite{jiang07}, that is attributed to interaction effects.

It has been debated in the literature
whether the Coulomb interactions in graphene
can be theoretically addressed within the standard Fermi-liquid-type
perturbation theory that requires, in particular that the particle
energy is much larger than the decay rate ($\Gamma\ll\epsilon$).
In order to understand to what extent graphene is a Fermi liquid,
one has to explore interaction-induced inelastic collision rates.
The inelastic quantum scattering rate at zero temperature (and finite
quasiparticle energy) has been considered in
Refs.~\onlinecite{Gonzalez96,Sarma07}. It was found in these works that the
behavior  at Dirac point is rather peculiar and requires a
careful incorporation of screening.

Another highly nontrivial feature of graphene is that inelastic
electron-electron collisions may limit the conductivity at the Dirac
point without any disorder or phonon
scattering \cite{Kashuba08,Fritz08,Foster09}.  This peculiarity of
graphene---which
should be contrasted to conventional systems where interactions do not
lead (in the absence of
Umklapp scattering) to finite resistivity---is a consequence of the
particle-hole symmetry
and decoupling between velocity  and momentum.
As a result, although the total momentum of interacting particles is conserved
during inelastic collisions, the total current may relax. The collision-limited
conductivity of undoped graphene is found to be inversely proportional
to $\alpha_g^2$ and depends on temperature
only through the renormalization of $\alpha_g$.
It was found that the energy relaxation caused by inelastic processes
in graphene is fast, which allows one to treat the
problem by using a relativistic hydrodynamic
approach \cite{Sachdev08,Mueller08,Mueller091,Mueller092}.

In this paper we analyze the inelastic electron-electron scattering in graphene
within the Keldysh diagrammatic approach. While we focus on the
equilibrium situation in this work, we have in mind to extend the
treatment of physics of inelastic processes
to strongly non-equilibrium setups, which explains why we
prefer to work in the framework of the Keldysh formalism.
More specifically, our main results are as follows:

\begin{itemize}

 \item We demonstrate that finite
   temperature strongly affects the screening properties of graphene,
   which, in turn, influences the inelastic scattering rates as
   compared to the zero-temperature case.

 \item Focussing on the high-temperature regime, we calculate the
   quantum scattering rate which
is relevant for dephasing of interference processes. We identify an
hierarchy of regimes
arising due to the interplay of a plasmon enhancement of the scattering
and finite-temperature
screening of the interaction. 

 \item We further discuss the energy relaxation and transport
   scattering rate in graphene.
Our results complement the kinetic-equation and hydrodynamic
approaches\cite{Kashuba08,Fritz08,Foster09}
for the collision-limited conductivity of clean graphene.

\end{itemize}

The paper is organized as follows. In section \ref{SecMod} we define
the model and develop
the Keldysh diagrammatic formalism for treating the problem of Coulomb
interaction in graphene.
Section \ref{SecPolopT} is devoted to the analysis of polarization
operator of graphene.
We first review and discuss the results for the zero temperature case.
Then we turn to the case of finite temperature and discuss the
properties of the polarization operator.
In particular, we compare the approximate form of the dynamically
screened interaction propagator  with the exact numerical results.
In Sec.~\ref{SecRates} we calculate the inelastic scattering rates
in the random phase approximation.
Here we use the finite-temperature polarization operator obtained in
Sec.~\ref{SecPolopT} to treat the problem analytically. We discuss
the asymptotics of the inelastic rates and
show that Dirac fermions show no Fermi liquid behavior. Further, in
Sec.~\ref{SecTrateanCond},
we discuss the collision-limited conductivity obtained within the
diagrammatic approach. In Sec.~\ref{SecConcl} we
conclude and summarize the main results of this paper.
Technical details of the calculations are presented in three Appendices.

\section{The Model}
\label{SecMod}
\subsection{Clean graphene with Coulomb interaction}
\label{SecGraphenanCoul}
In this paper we consider clean graphene near the degeneracy point.
The problem is described by the following Hamiltonian, which is a sum of
the Dirac Hamiltonian $\hat{H}_0$ (describing the physics of
non-interacting electrons in graphene at not too high energies)
and the Coulomb interaction term,

\begin{widetext}
\begin{equation}\label{GlModel}
 \hat{H}=\hat{H}_0+\hat{V} =\! \sum\limits_{\nu}
\int d^2r
{}{}\,
 \fkt{\hat{\Psi}^{\dagger}_{\nu}}{\vec{r}}
 \fkt{}{-iv_F\boldsymbol{\sigma}\!\cdot\!\vec{\nabla}}
 \fkt{\hat{\Psi}_{\nu}}{\vec{r}} + \frac{1}{2}\sum\limits_{\nu ,\nu '}
\int d^2r_1 d^2r_2
{}{}\,
\fkt{\hat{\Psi}^{\dagger}_{\nu}}{\vec{r}_1}\fkt{\hat{\Psi}^{\dagger}_{\nu
     '}}{\vec{r}_2}\frac{e^2}
{\varepsilon |\vec{r}_1-\vec{r}_2|}
 \fkt{\hat{\Psi}_{\nu '}}{\vec{r}_2}\fkt{\hat{\Psi}_{\nu}}{\vec{r}_1}.
\end{equation}
\end{widetext}
Here $\varepsilon$ is the dielectric constant. The spinors $\hat{\Psi}$ have two components in the sublattice
space, $\sigma_i$ are Pauli matrices operating in this space.
The indices $\nu,\nu'$ label  $N$ independent
degrees of freedom (in graphene $N=4$ accounts for spin and valleys degeneracy):  the Coulomb interaction is invariant
with respect to any rotations in the corresponding space. We set $\hbar=1$.
We focus on the case of undoped graphene and set the chemical potential
(counted from the Dirac point) to zero, $\mu=0$.

The retarded (advanced) Green's function of the noninteracting
Hamiltonian $\hat{H}_0$
(the bare Green's function) in the energy-momentum space has the form
\begin{equation} \label{GlGreenschefktretav}
 \fkt{G^{R,A}_0}{\epsilon,\vec{p}} = \frac{\epsilon\mathds{1}
   +v_F\boldsymbol{\sigma}\cdot\vec{p}}{(\epsilon\pm i0)^2-v_F^2p^2}\,.
\end{equation}

It is convenient to introduce the projection operators that distinguish between the
two chiral states:
\begin{equation}
 \mathcal{P}_{\pm}(\vec{p})=\frac{\mathds{1}\pm \boldsymbol{\sigma}\cdot\vec{n}}{2},
\label{projectors}
\end{equation}
where $\vec{n}_p=\vec{p}/p$ is the unit vector in the direction of momentum.
With the help of Eq.~\eqref{projectors} the matrix Green's function, Eq.~\eqref{GlGreenschefktretav},
can be decomposed into the superposition of the two Green's functions corresponding to the states
with $+$ and $-$ chiralities:
\begin{equation} \label{GreenProj}
 \fkt{G^{R,A}_0}{\epsilon,\vec{p}} = \mathcal{P}_{+}(\vec{p})\fkt{G^{R,A}_{0+}}{\epsilon,\vec{p}}
+\mathcal{P}_{-}(\vec{p})\fkt{G^{R,A}_{0-}}{\epsilon,\vec{p}},
\end{equation}
where
\begin{equation}
 \fkt{G^{R}_{0\pm}}{\epsilon,\vec{p}}=\frac{1}{\epsilon+i0\mp v_F p}
\label{chiralG}
\end{equation}
and $\fkt{G^{A}_{0\pm}}{\epsilon,\vec{p}}=\left[ \fkt{G^{R}_{0\pm}}{\epsilon,\vec{p}}\right]^*$.
For the later purposes we will need the quasiparticle spectral weight
\begin{multline} \label{GlSpektralesgewicht}
 \mathcal{A}_0(\epsilon,\vec{p})=\frac{1}{2i}\left[\fkt{G^{R}_0}{\epsilon,\vec{p}}
-\fkt{G^{A}_0}{\epsilon,\vec{p}}\right]=\\
 -\frac{\pi}{2\epsilon}
 \fkt{}{\epsilon\mathds{1}+v_F\boldsymbol{\sigma}\cdot\vec{p}}
\fktb{}{\fkt{\delta}{\epsilon-v_F
       p}+\fkt{\delta}{\epsilon+v_F
       p}}.
\end{multline}
Using the projection operators, we decompose the spectral weight as follows:
\begin{eqnarray} \label{ProjSpektralesgewicht}
 \mathcal{A}_0(\epsilon,\vec{p})
&=&\mathcal{P}_{+}(\vec{p})\mathcal{A}_{0+}(\epsilon,\vec{p})+\mathcal{P}_{-}(\vec{p})\mathcal{A}_{0-}(\epsilon,\vec{p}),\\
\mathcal{A}_{0\pm}(\epsilon,\vec{p})&=&-\pi\delta(\epsilon\mp v_F p).
\end{eqnarray}
It is worth noticing that for Dirac particles the spectral weight $\mathcal{A}_0(\epsilon,\vec{p})$
is not given by the imaginary part of the Green's function $\fkt{G^{R}_{0}}{\epsilon,\vec{p}}$, because
the latter contains the Pauli matrix $\sigma_y$. However, within each chirality the conventional relation holds:
$\mathcal{A}_{0\pm}(\epsilon,\vec{p})=\Im\, \fkt{G^{R}_{0\pm}}{\epsilon,\vec{p}}.$

Next, we introduce the coupling constant for Coulomb interaction in graphene
\begin{equation}
 \alpha_g=\frac{e^2}{\varepsilon v_F},
\label{alphag-def}
\end{equation}
which is similar to the fine structure constant
$$\alpha= \frac{e^2}{c}\approx \frac{1}{137}$$
but is $c/\varepsilon v_F$ times
larger.  The bare propagator of the Coulomb interaction in
\eqref{GlModel} reads (in momentum space):
\begin{equation} \label{EqInteraction}
\fkt{D_0}{\vec{q}}=\frac{2\pi \alpha_g v_F }{|\vec{q}|}
\end{equation}
Throughout the paper we assume $\alpha_g\ll 1$. This assumption is favored by recent experiments
~\cite{kozikov10,Reed10} which suggested that the effective interaction constant in 
graphene is rather small at experimentally relevant temperatures.

\subsection{Keldysh formalism}
\label{SecDescBasOP}
Although in this work we discuss only equilibrium physics, we use the Keldysh
formalism\cite{Kamenev09,Smith86} in order to have a basis which can
be generalized to the nonequilibrium
situation. In particular, this Keldysh formalism will be used
elsewhere for deriving kinetic equations
for clean and disordered graphene.

The bare Green's function is now a matrix in the Keldysh space
\begin{equation}\label{EqKeldyshGf}
 \check{G}_0=\begin{pmatrix}G^R_0&G^K_0\\0&G^A_0\end{pmatrix}.
\end{equation}
Here the Keldysh component $G^K_0$ at the equilibrium reads
\begin{multline}\label{EqKeldyshComp}
 \fkt{G^K_0}{\epsilon,\vec{p}}=f(\epsilon)
\fktb{}{\fkt{G^R_0}{\epsilon,\vec{p}}
-\fkt{G^A_0}{\epsilon,\vec{p}}}\\
=2i f(\epsilon) \mathcal{A}_0(\epsilon,\vec{p}),
\end{multline}
where the fermionic thermal factor is given by $\fkt{f}{\epsilon}=\fkt{\tanh}{\epsilon/2T}$.

The full Keldysh Green's function for Eq.~\eqref{GlModel} is expressed
through $\check{G}_0$
\begin{equation} \label{EqfullGf}
  \check{G} = \fkt{}{\check{G}_0-\check{\Sigma}}^{-1}
\end{equation}
where $\Sigma$ is the full self energy.
In the lowest order in the fully dressed propagator of Coulomb interaction $
\check{D}$ the retarded self-energy is given by
\begin{equation} \label{EqleadingselfEn}
 \Sigma^R_0=\frac{i}{2}\fkt{}{D^K\circ G^R_0+D^R\circ G^K_0}\,,
\end{equation}
where the symbol $\circ$ denotes integration over all
internal energies and momenta.

In the equilibrium situation the Keldysh component $D^K$ of the
interaction propagator satisfies
\begin{equation} \label{EqKeldyshInter}
 \fkt{D^K}{\omega,\vec{q}}=2 i
\fkt{g}{\omega}
\Im \fkt{D^R}{\omega,\vec{q}}\,,
\end{equation}
where $\fkt{g}{\omega}=\fkt{\coth}{\omega/2T}$ is the bosonic thermal
factor. Using  Eqs.~\eqref{EqleadingselfEn}, \eqref{EqKeldyshInter},
and \eqref{EqKeldyshComp}, one gets
\begin{equation}
\label{EqImSelfEnDef}
\Sigma^R_0-\Sigma_0^A=-2i \left[\fkt{}{f+g}\Im\, D^R\right] \circ (G^R-G^A)\,.
\end{equation}
The retarded self energy is a matrix which contains the two terms:
\begin{equation}
 \Sigma^R=\Sigma^R_\epsilon\mathds{1}+\Sigma_v^R\, \boldsymbol{\sigma}\cdot\vec{n}_p.
\end{equation}
The real parts of $\Sigma^R_\epsilon$ and $\Sigma_v^R$ give rise to the corrections to
the energy and Fermi velocity, respectively:
\begin{eqnarray}
\delta\epsilon&=&-\Re\Sigma^R_\epsilon(\epsilon,\vec{p}),\\
\delta v_F&=&\Re\Sigma^R_{v}(\epsilon,\vec{p})/p.
\end{eqnarray}
In order to fix the energy unrenormalized, we introduced the $Z$ factor
\begin{equation}
 Z(\epsilon,\vec{p})=1-\frac{\Re\Sigma^R_\epsilon(\epsilon,\vec{p})}{\epsilon},
\label{Zfactor}
\end{equation}
so that $\epsilon-\Re\Sigma^R_\epsilon(\epsilon,\vec{p})=Z(\epsilon,\vec{p})\epsilon$.
The renormalized Fermi velocity takes then the form
\begin{equation}
 v_F^*=v_F Z(\epsilon,\vec{p})\left[1+\frac{\Re\Sigma^R_{v}(\epsilon,\vec{p})}{v_F p}\right].
\label{vFstar}
\end{equation}

Using Eqs.~\eqref{Zfactor} and \eqref{vFstar}, we write the retarded Green's function as
\begin{multline}
\label{GRself}
 \fkt{G^R}{\epsilon,\vec{p}} \\ =
Z\left[(\epsilon-iZ\Im\Sigma_\epsilon^R)\mathds{1}
-\left(v_F^*+\frac{iZ\Im\Sigma_v^R}{p}\right)
\boldsymbol{\sigma}\cdot\vec{p}\,\right]^{-1},
\end{multline}
where
\begin{eqnarray}
 \Im\Sigma_\epsilon^R&=&-\frac{1}{2}\Im D^R(f+g)\circ \mathrm{Tr}\, \mathcal{A}, \label{ImSigmaE}\\
\Im\Sigma_v^R&=&-\frac{1}{2}\Im D^R(f+g)\circ \mathrm{Tr}\,\mathcal{A}\, \boldsymbol{\sigma}\cdot\vec{n}_p.
\label{ImSigmav}
\end{eqnarray}
Note that  in this representation the velocity acquires a non-zero imaginary part.

The most singular terms in the renormalized velocity $v_F^*$, coupling constant $\alpha_g^*=e^2/\varepsilon v_F^*$, and 
the $Z$-factor can be summed up
by means of the renormalization group approach~\cite{Gonzalez94}. For the case of weak interaction $\alpha_g\ll 1$
the solution of one-loop RG equations has the form:
\begin{eqnarray}
v_F^*(\epsilon)&=&v_F\left(1+\frac{\alpha_g}{4}\ln\frac{\Lambda}{\epsilon}\right),
\label{vstar}\\
\alpha_g^*(\epsilon)&=&\frac{\alpha_g}{1+\frac{\alpha_g}{4}\ln\frac{\Lambda}{\epsilon}},\\
Z(\epsilon)&=&\exp\left\{-\frac{4}{3\pi}\left[\alpha_g-\alpha_g^*(\epsilon)\right]\right\}\nonumber \\
&=&\exp\left[-\frac{1}{3\pi}\frac{\alpha_g^2\ln\frac{\Lambda}{\epsilon}}{1+\frac{\alpha_g}{4}\ln\frac{\Lambda}{\epsilon}} \right]
\simeq 1.
\end{eqnarray}
Here $\Lambda$ is the ultraviolet energy cutoff (bandwidth) and the on-shell relation between $\epsilon$ and $p$ 
is assumed. At finite temperature $T$ the renormalization stops at $\mathrm{max}[\epsilon,T]$. Therefore, for energies
below $T$ (which will be in the focus below) the renormalized velocity and the $Z$-factor are independent of energy.
Since for $\alpha_g\ll 1$ the corrections to unity in the renormalized $Z$-factor are parametrically small,
in what follows we set $Z=1$.

Similarly to the bare Green's function, Eq.~\eqref{GreenProj}, the full Green's function, Eq.~\eqref{GRself}, can be
represented as a sum of the two terms corresponding to $\pm$-chiralities:
\begin{equation} \label{GreenFullProj}
 \fkt{G^{R}}{\epsilon,\vec{p}} = \mathcal{P}_{+}(\vec{p})\fkt{G^{R}_{+}}{\epsilon,\vec{p}}
+\mathcal{P}_{-}(\vec{p})\fkt{G^{R}_{-}}{\epsilon,\vec{p}},
\end{equation}
where
\begin{equation}
 \fkt{G^{R}_{\pm}}{\epsilon,\vec{p}}=\frac{Z}{\epsilon \mp v_F^* p-i Z \Im \Sigma_{\pm}^R}
\end{equation}
with
\begin{multline}
 \Im \Sigma_{\pm}^R = \Im\Sigma_\epsilon^R\pm \Im\Sigma_v^R \\
 = -\Im D^R(f+g)\circ \mathrm{Tr}\left[\mathcal{A}\mathcal{P}_{\pm}\right].
\label{ImSigmaPM}
\end{multline}

Clearly, the bare Coulomb interaction (whose propagator
is purely real) 
does not yield the imaginary part of the self-energy, so that one has to
take into account the retardation effects.
In the random phase approximation (RPA), the screened Coulomb
interaction takes the form
\begin{equation}\label{EqRPAintDef}
 \fkt{D^R_{\mathrm{RPA}}}{\omega,\vec{q}}
=\frac{\fkt{D_0}{q}}{1+\fkt{D_0}{q}N\fkt{\Pi^R}{\omega,\vec{q}}}
\end{equation}
where $N$ is the number of flavors.
The dynamical screening in Eq. \eqref{EqRPAintDef} is expressed through the bare
polarization operator
\begin{equation} \label{EqPolopDef}
 \Pi^R=\frac{i}{2}\fkt{\mathrm{Tr}}{G^R_0\circ G^K_0+G^K_0\circ G^A_0}.
\end{equation}
In Sec. \ref{SecPolopT} below we study $\Pi$ at zero and finite temperature.

\section{Polarization Operator}
\label{SecPolopT}

In the present section we discuss the properties of the
  polarization operator. This is of primary importance for understanding the
  screening of the electron-electron interaction and thus the physics
  of interaction-induced phenomena. We find it instructive to start with
  analyzing the zero temperature result and describing the
  processes relevant for the polarization operator. Then we turn to the
  case of finite temperature, which is our main interest in the paper.
Finally, we analyze the consequences for the RPA-screened interaction.
In the end of this section we comment on the applicability of the RPA in graphene.

\subsection{Polarization Operator at Zero Temperature and RPA
  Interaction Propagator}

\label{SecPolopzT}

The polarization operator in the energy-momentum representation reads
\begin{multline}\label{EqPolopstartCalc}
 \fkt{\Pi^R}{\omega,\vec{q}}
=-\pintd{p}{2}{}{}\pint{\epsilon}{}{}
\fkt{f}{\epsilon}\fktc{\mathrm{Tr}}{\fkt{\mathcal{A}_0}{\epsilon,\vec{p}}\right.\\
\times \left.\fktb{}
{\fkt{G^R_0}{\epsilon+\omega,\vec{p}+\vec{q}}
+\fkt{G^A_0}{\epsilon-\omega,\vec{p}-\vec{q}}}}\,,
\end{multline}
where $G^{R,A}_0$ are bare Green functions
\eqref{GlGreenschefktretav}.
The momentum integrals that appear in this expression can be
conveniently evaluated using
elliptic coordinates as described in Appendix \ref{AppPolarisation-Operator}.

In the zero temperature limit, Eq.~\eqref{GlPolopSimplifyedIm}
simplifies and leads to
the well-known expression for the
imaginary part of the polarization operator:\cite{Gonzalez94}
\begin{equation} \label{EqZeroTempImPolop}
 \fkt{\Im\Pi^{R}}{\omega,\vec{q}}
=\frac{1}{16}\Re\frac{q^2 \mathrm{sign}\omega}{\sqrt{\omega^2-v_F^2q^2}}.
\end{equation}
One can see that the imaginary part of the polarization operator is
non-vanishing only if $\omega>v_F q$ and shows a
divergence at the ``light cone'' $\omega=v_F q$.

\begin{figure}
[b]
\begin{center}
 \includegraphics[width=3.7cm]{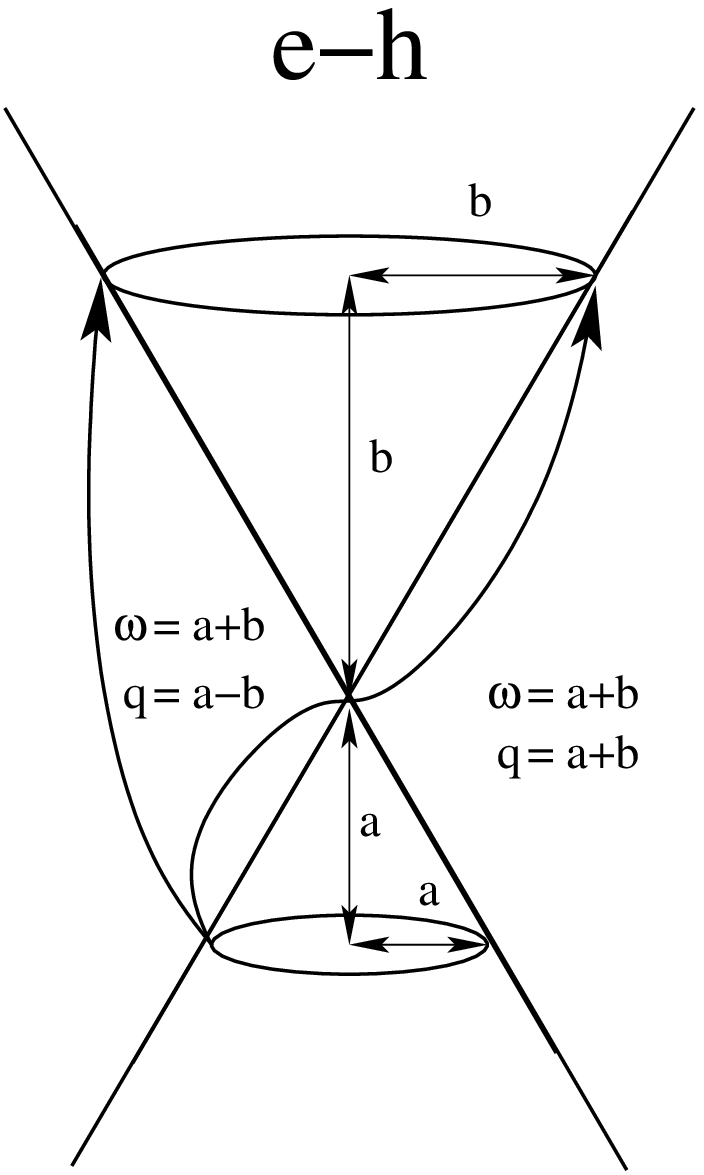} \hspace*{1cm}
 \includegraphics[width=2.9cm]{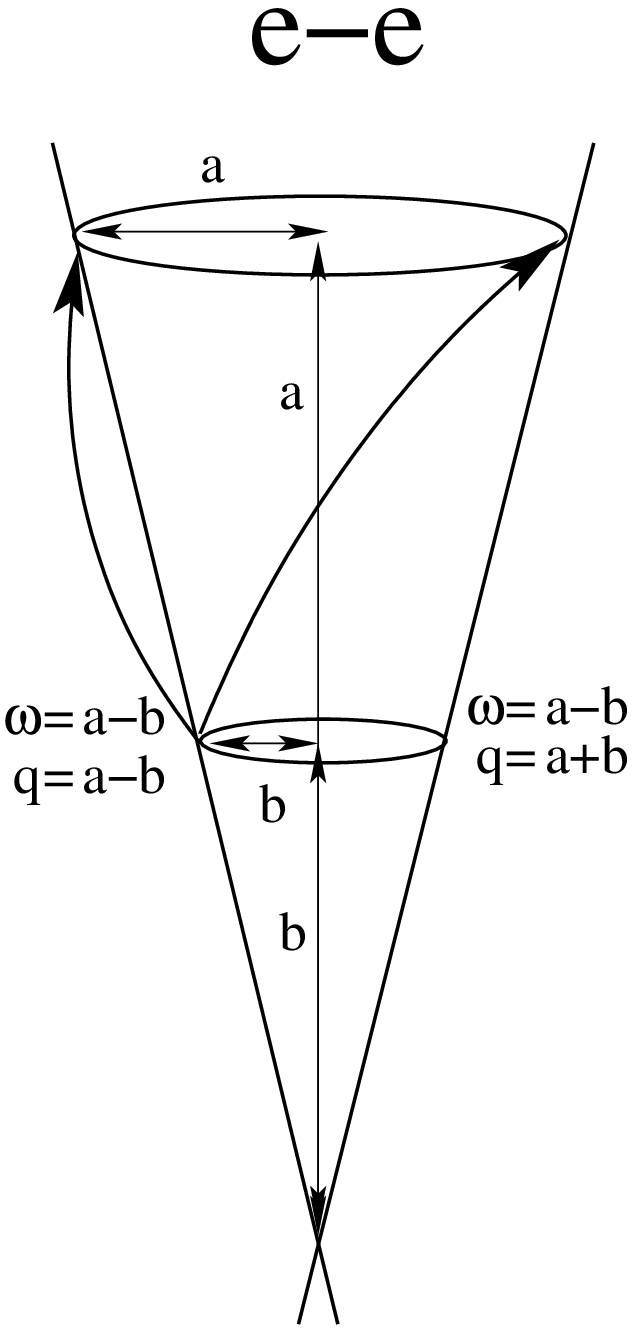}
\caption{ Schematics of electron-hole (e-h) creation (left panel) vs electron-electron (e-e)
  scattering (right panel) near the Dirac point for $\mu=0$. The hole-hole scattering
  processes are analogous to those shown in the right panel.  We set $v_F=1$ for brevity. In both panels the possible transferred
  momenta satisfy $|a-b|\leq |\vec{q}|\leq a+b$ (only the extreme cases of minimal and maximal $q$ are shown).
  For e-h creation $\omega=a+b$, so that $\omega\geq qv_F$, whereas for e-e scattering the kinematic restrictions
  yield $\omega=a-b\leq qv_F$.
  }
\label{FigPolScetch}
\end{center}
\end{figure}

To understand the vanishing of the imaginary part of the polarization operator
at $\omega<v_F q$, it is instructive to analyze
\cite{Gonzalez96,Foster09} the kinematic restrictions for elementary
processes.  It is easy to see that
an on-shell electron-hole pair can be created if
(left panel of Fig.~\ref{FigPolScetch})
\begin{equation}\label{EqCondforPHcreation}
 \omega \geq v_F q \,.
\end{equation}
This should be contrasted to the energy and momentum conservation of on-shell
electron-electron scattering processes
(right panel of Fig.~\ref{FigPolScetch})
\begin{equation} \label{EqEnMomConsofEl}
 \omega^2+2v_F^2p_1p_2\fkt{}{1-\fkt{\cos}
{\fktb{\sphericalangle}{\vec{p}_1,\vec{p}_2}}}=v_F^2q^2\,,
\end{equation}
which implies the condition
\begin{equation}\label{EqCondforEEscatt}
 \omega \leq v_F q.
\end{equation}

Thus, if we restrict our consideration to on-shell particles,
electron-electron scattering processes only create electron-hole pairs
under the condition $\omega=v_F q$, when $\Im\Pi$ diverges.
As follows from Eq.~\eqref{EqEnMomConsofEl}, scattering
processes that satisfy $\omega=v_F q$ correspond to forward scattering with
$\fktb{\sphericalangle}{p_1,p_2}=0$.
After the RPA resummation, the interaction propagator takes the form
\begin{equation} \label{EqZeroTRPAInt}
 \fkt{D^R_{\mathrm{RPA}}}{\omega,q}=\frac{\displaystyle 2\pi v_F
   \alpha_g}{\displaystyle q+\frac{\pi v_F \alpha_g N }{8}\frac{i
     q^2}{\sqrt{\omega^2-v_F^2q^2}}}.
\end{equation}
It is seen that the divergence at $\omega=v_F q$  is now eliminated; moreover,
the imaginary part of the propagator is zero at the light cone.
Therefore, the forward scattering divergence that arises on the Golden Rule (GR)
level disappears within RPA, yielding a zero scattering rate
on the RPA level. This is a manifestation of a highly singular
character of the zero-temperature problem where RPA may be
insufficient. We will see below that at finite temperature the thermal
broadening regularizes the problem, so that scattering rates can be
evaluated within the RPA.

\begin{figure}[t]
 \includegraphics[width=0.75\columnwidth]{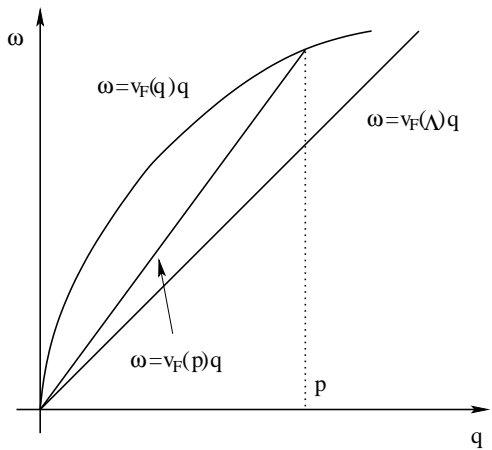}
\caption{ Interaction-induced dispersion
  correction near the Dirac cone. The electron-hole creation
  [$\omega\geq \fkt{v_F}{q}q$] and
electron-electron scattering
[$\omega\leq\fkt{v_F}{p}q$] regions do not overlap, so that the
zero-temperature inelastic scattering rate is zero. }
\label{FigLogCorSceme}
\end{figure}

In view of the singular character of the problem,
at zero temperature the imaginary part of the self-energy
is highly sensitive to changes of the electron dispersion~\cite{Foster09}.
The Fermi velocity depends logarithmically on the momentum or energy~\cite{Gonzalez94},
Eq.~\eqref{vstar},
due to the renormalization by Coulomb interaction. In
Eq.~\eqref{EqZeroTempImPolop} we have neglected
the momentum dependence of $v_F$ which
leads to a separation
of the two regions defined by Eqs. \eqref{EqCondforPHcreation} and
\eqref{EqCondforEEscatt},
see Fig.~\ref{FigLogCorSceme}.
The electron-hole creation [$\omega\geq \fkt{v_F}{q}q$] and
electron-electron scattering
[$\omega\leq\fkt{v_F}{p}q$] regions are then separated
due to the
renormalization-induced nonlinearity of the electron dispersion, which
leads to the vanishing
of the zero-$T$ scattering rate already at the GR level.

However, at finite temperature the situation is essentially
different.
First, the conditions
\eqref{EqCondforPHcreation} and \eqref{EqCondforEEscatt}
will be smeared by temperature.
Second,
for energy scales smaller than $T$ the renormalization of Fermi
velocity is cut off by
temperature and hence, the linearity of the dispersion relations is
restored: renormalization reduces merely to a
$T$-dependence of the Fermi velocity.
Therefore, when discussing the finite-$T$ physics of inelastic
scattering on scales $\alt T$ we will disregard the
renormalization-induced nonlinearity of the dispersion.
Furthermore, below, whenever we will use the notation $v_F$ for $\epsilon\alt T$ we
will mean the renormalized value of the velocity $v_F^*(T)$ and omit
the asterisk for brevity.

\subsection{Polarization Operator at non-Zero Temperature}
\label{SecPolopfT}

We are now ready to calculate the polarization operator at finite temperature.
It turns out that the effect of finite temperature on the screening in
graphene is much more pronounced than
in conventional metals with finite Fermi-surface and quadratic electronic dispersion. Indeed,
the linearity of the spectrum of
Dirac fermions gives rise to a strong (linear) energy dependence of
the density of states in two dimensions,
whereas for parabolic spectrum the density of states is constant. In
the latter case, the polarization operator
is essentially independent of temperature. This is not the case for
Dirac particles.

\begin{figure}[b]
 \includegraphics[width=\columnwidth]{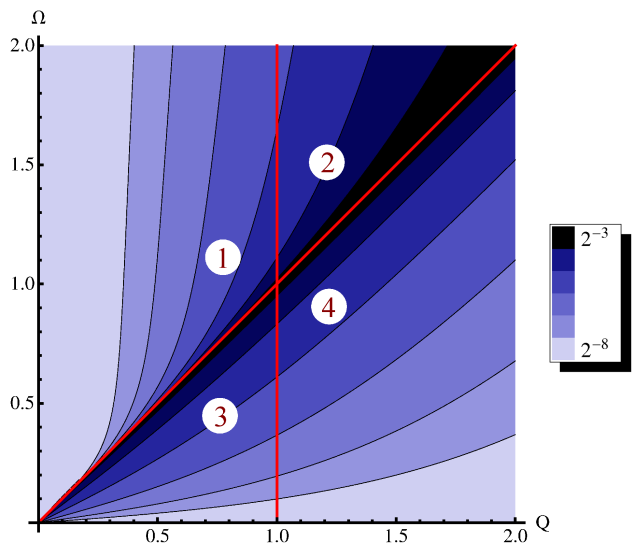}
\caption{(Color online) Imaginary part of the polarization operator in the
  frequency-momentum plane. Four regions of dimensionless variables
  $\Omega$ and $Q$ (see text) are indicated.}
\label{FigFinPolopAreas}
\end{figure}

Physically, a finite temperature leads to population of electronic states
in an energy range $\sim T$ around the Fermi level.
Let us consider the undoped graphene where the chemical potential
lies at the Dirac point.
The typical density of states that participate in the screening of
Coulomb interaction
at finite temperature is now proportional to $T$.
In more technical terms, at finite $T$ the
integration over the fermionic energy
in the polarization bubble essentially
involves not only the distribution function [thermal factors
$f(\epsilon)$] but also the Green's functions.
This strongly changes the polarization
operator at $qv_F,\omega\alt T$.

To simplify the notations at non-zero temperature, we introduce the
dimensionless variables according to
\begin{eqnarray}
\label{EqDimLesCoordinates}
 Q=\frac{v_Fq}{2T}\ , \qquad && \Omega=\frac{\omega}{2T}.
\end{eqnarray}
We use the general expression for the polarization operator,
Eqs.~\eqref{GlPolopSimplifyedIm} and \eqref{GlPolopSimplifyedRe}
from Appendix \ref{AppPolarisation-Operator}.
Considering four different regions shown in Fig.~\ref{FigFinPolopAreas},
we  simplify these equations in each of the cases, which allows us
to treat the problem analytically.

The four regions are defined by the following conditions:
\begin{itemize}
 \item Region 1: $Q\ll 1$ and $Q<|\Omega|$
 \item Region 2: $1\ll Q$ and $Q<|\Omega|$
 \item Region 3: $Q\ll 1$ and $|\Omega|<Q$
 \item Region 4: $1\ll Q$ and $|\Omega|<Q$
\end{itemize}

As shown in Appendix \ref{AppPolarisation-Operator}, the leading-order
terms with non-vanishing
imaginary part form the following simplified polarization operator:
\begin{widetext}
\begin{equation}\label{EqSimpPolopatnonZT}
 \fkt{\Pi^{R}}{\Omega,Q}=\frac{T}{v_F^2}
 \left\{\begin{array}{lr}\begin{array}{lcc}
\displaystyle{\frac{\ln 2}{\pi}\fktb{}{1-\frac{\ABS{\Omega}}
{\sqrt{\Omega^2-Q^2}}}
+\frac{i}{8}\frac{Q^2\fkt{\tanh}
{\Omega/2}}{\sqrt{\Omega^2-Q^2}}, } &\quad (\mathrm{region}\ 1)&\quad  |\Omega|>Q \\
\displaystyle{\frac{\ln 2}{\pi}+ \frac{i\ln 2}{\pi}
\frac{\Omega}{\sqrt{Q^2-\Omega^2}}, }&\quad (\mathrm{region}\ 3)&\quad |\Omega|<Q
			\end{array} & Q\ll 1 \\
			\begin{array}{lcc}
\displaystyle{\frac{i}{8}\frac{Q^2\, \mathrm{sign}\Omega}{\sqrt{\Omega^2-Q^2}}, }&\quad(\mathrm{region}\ 2)&\quad |\Omega|>Q\\
\displaystyle{\frac{1}{8}\frac{Q^2}{\sqrt{Q^2-\Omega^2}}+\frac{i}{\sqrt{2\pi}}
\frac{\sqrt{Q}e^{-Q}\sinh\Omega}
{\sqrt{Q^2-\Omega^2}}, }\hspace*{0.73cm}&\quad (\mathrm{region}\ 4)&\quad |\Omega|<Q
						\end{array} & Q\gg 1
		    \end{array}\right.
\end{equation}
\end{widetext}
In the case of large momenta, $Q\gg 1,$ we recover the
zero-temperature result.
For $Q\ll 1$, the polarization operator substantally differs from the
zero-$T$ expression.
We will discuss this case in more detail in section \ref{SecRPAIntatnonZero}.
The separation between $\Omega<Q$ and $\Omega>Q$ is dictated by the non-analytical  structure
of the polarization operator, see Eq.~\eqref{EqSimpPolopatnonZT}.

\subsection{RPA Interaction at non-Zero Temperature}
\label{SecRPAIntatnonZero}

The real part of the polarization operator,
Eq.~\eqref{EqSimpPolopatnonZT}, for $Q\ll 1$
is determined by temperature and leads therefore to the screening of
Coulomb interaction: 
\begin{multline} \label{EqScreeninglength}
  \lim_{\substack{\Omega\rightarrow 0}}
\fkt{D^R_{\mathrm{RPA}}}{\Omega,Q}=\frac{\fkt{D_0}{Q}}{1+\fkt{D_0}{Q}N\fkt{\Pi^R}{0,Q}}\\
= \frac{v_F^2\pi}{NT\ln 2}\left(\frac{\alpha_g N \ln 2}{Q+\alpha_g N \ln 2}\right),
\end{multline}
which yields the
screening length 
\begin{equation}
l_{\mathrm{scr}}= \frac{v_F}{2 \alpha_g N T \ln 2}.
\end{equation}
Thus at finite temperature the system does screen the long range
Coulomb interaction. Note that for $Q\ll \alpha_g N$, the RPA propagator
becomes independent of the interaction constant $\alpha_g$,
\begin{equation}
 \left.D^R_{\mathrm{RPA}}(\Omega,Q)\right|_{Q\ll \alpha_g N}\simeq \frac{1}{\Pi^{R}(\Omega,Q)},
\end{equation}
as in conventional systems with Coulomb interaction. We will see, however, that the dominant contributions
to relaxation rates are determined by higher transferred momenta $Q\agt \alpha_g N$, 
where the peculiarities of the finite-$T$ screening in graphene are crucially important.

In Region 1, the real part of the polarization operator in
Eq.~\eqref{EqSimpPolopatnonZT} may become negative, leading to
emergence of plasmon excitations.
The plasmon dispersion $\Omega_p(Q)$ is determined by the zero of
\begin{equation} \label{EqPlasmonDisp}
 1+D_0N\Pi^R\!=\!
\frac{\alpha_g N Q \ln 2}
{\fkt{}{\Omega^2-Q^2}^{3/2}}
\fktb{}{\!\Omega\!-\!
\frac{\fkt{}{\alpha_gN\ln 2\!+\!Q} \sqrt{Q}}{\sqrt{Q+\alpha_gN
       2 \ln 2}}\!},
\end{equation}
yielding
\begin{equation}
\fkt{\Omega_p}{Q}=
\frac{\fkt{}{\alpha_gN\ln 2+Q}\sqrt{Q}}{\sqrt{Q+2\alpha_gN
      \ln 2}}.
\end{equation}
For $Q\to 0$, this simplifies to $\Omega_p(Q)\propto \sqrt{\alpha_g N Q}$.
A non-zero imaginary part of the polarization operator
\eqref{EqSimpPolopatnonZT} in the corresponding region implies that
these plasmons have a finite lifetime.
The decay rate of plasmon excitations is given by
\begin{equation}
\fkt{\Gamma_p}{\Omega,Q}=\frac{\fkt{}{\Omega^2-Q^2}^{3/2}}{\alpha_gN Q \ln 2} \Im \fkt{D_0}{Q}\fkt{\Im \Pi^R}{\Omega,Q},
\end{equation}
which yields
\begin{multline}\label{EqPlasmonWidth}
 \left.\fkt{\Gamma_p}{\Omega,Q}\right|_{\Omega=\fkt{\Omega_p}{Q}}=\left.\frac{\pi
     \Omega}{16\ln 2}\fkt{}{\Omega^2-Q^2}\right|_{\Omega=\fkt{\Omega_p}{Q}}
 \\ \leq \left.\frac{\pi
     \Omega^3}{16\ln 2}\right|_{\Omega=\fkt{\Omega_p}{Q}}\,.
\end{multline}
Remarkably, Eq.~\eqref{EqPlasmonWidth}
indicates a good quasiparticle behavior
for plasmons which is, as we will see in Sec.~\ref{SecQscatt}, not
true for electronic excitations.
The situation is somewhat similar to Luttinger liquid
where plasmons are almost perfect quasiparticles, whereas, from the
point of view of fermionic excitations, the Luttinger liquid represents a paradigmatic
example of a non-Fermi-liquid.

\begin{figure*}
 \showgraph{width=10cm}{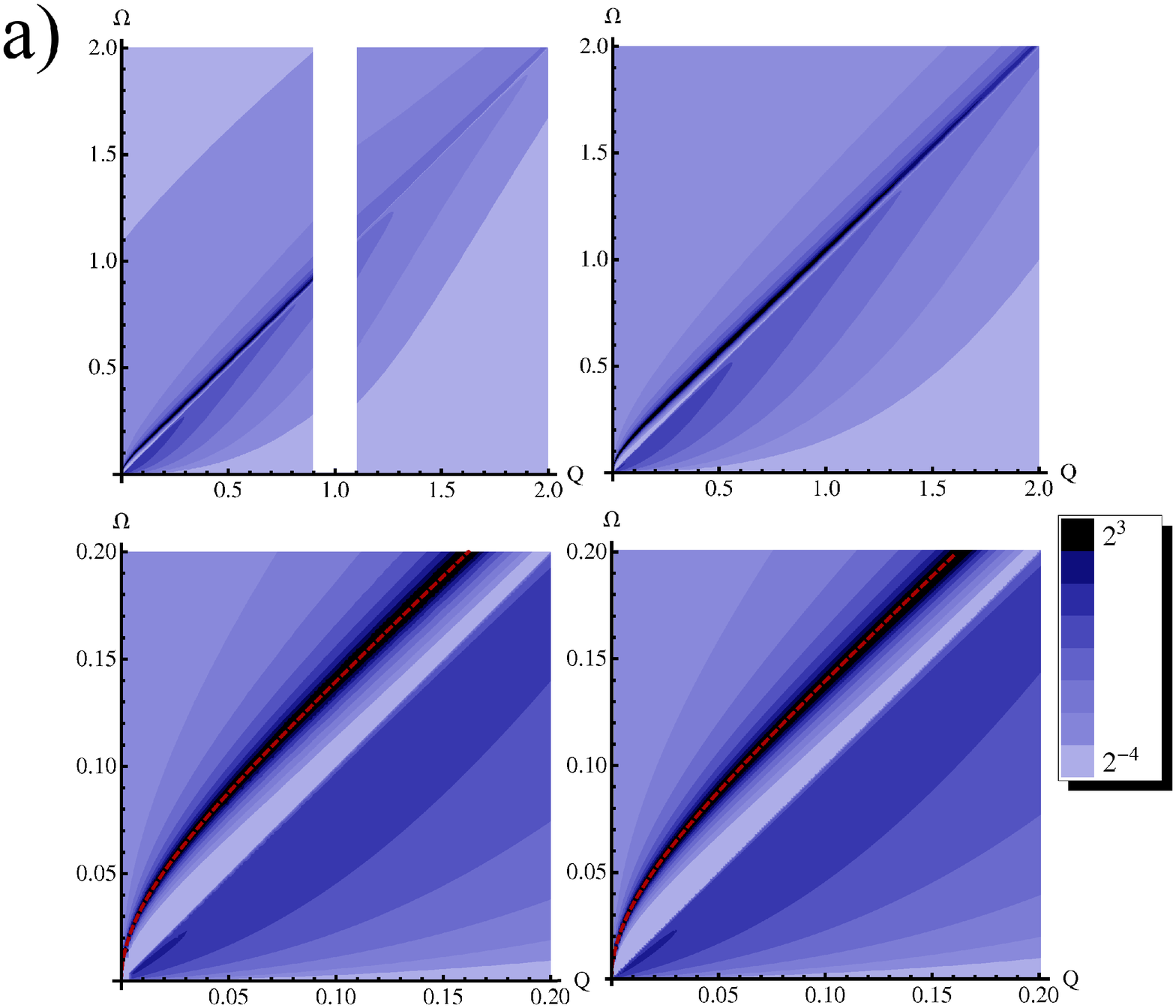}
 \showgraph{width=5cm}{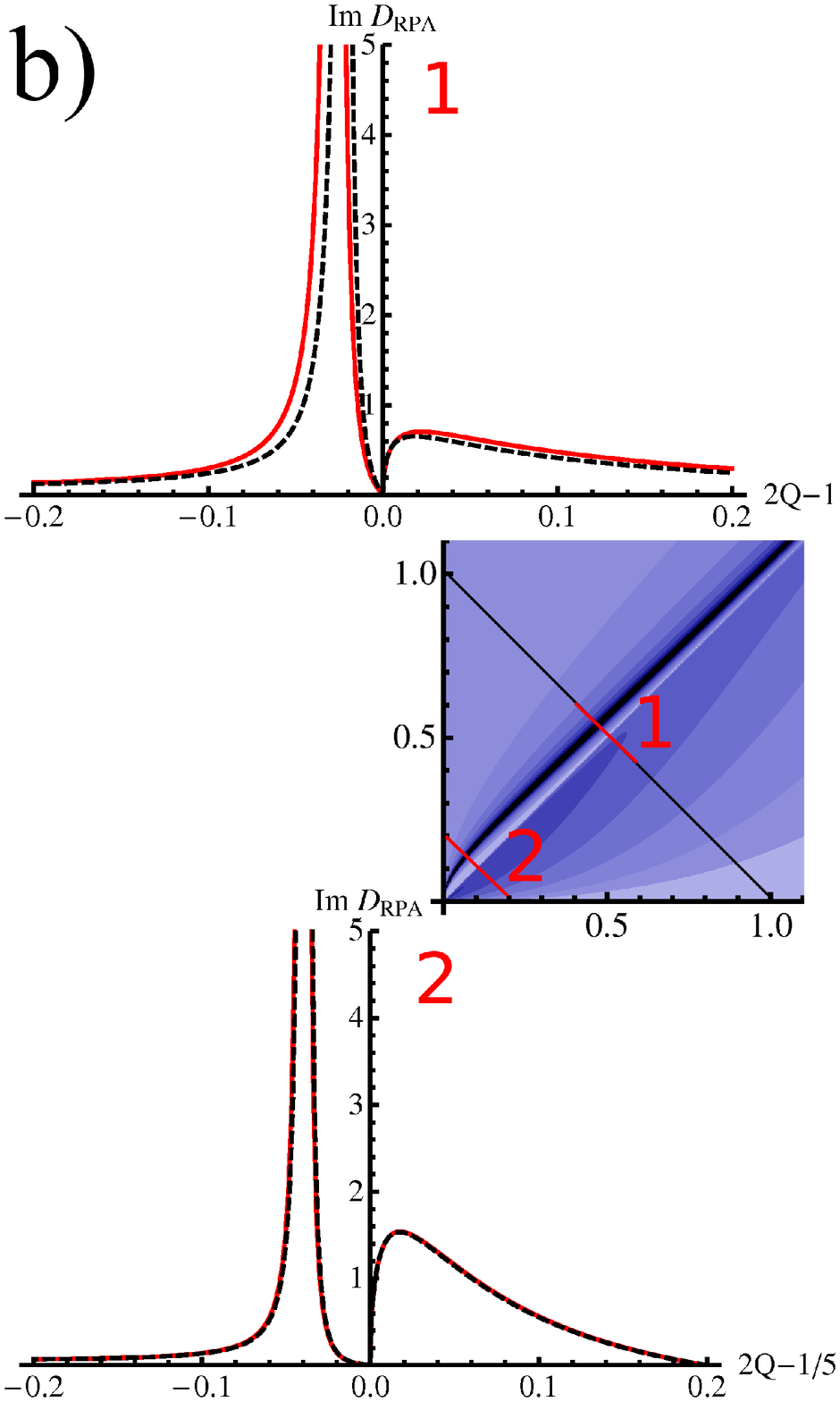}
\caption{(Color online) \textbf{a)} Imaginary part of RPA interaction. Left:
using an analytic approximation, right: by numerical evaluation of
Eqs.~\eqref{GlPolopSimplifyedIm} and \eqref{GlPolopSimplifyedRe}.
Dashed lines indicate the plasmon dispersion (see Eq.~\eqref{EqPlasmonDisp}).
\textbf{b)} crosssections through the plots of a)
from $(\Omega,Q)=(1,0)$ to $(\Omega,Q)=(0,1)$  (1)
from $(\Omega,Q)=(0.2,0)$ to $(\Omega,Q)=(0,0.2)$  (2).
Solid curves:
analytic approximation; dashed curves: numerical integration.
For all plots the interaction
strength is $\alpha_gN= 1/3$.  }
\label{FigIDRPA}
\end{figure*}
Figure \ref{FigIDRPA} demonstrates that Eq.~\eqref{EqSimpPolopatnonZT} for the
polarization operator yields a remarkably good approximation for
evaluation of the imaginary part of the RPA-interaction,
\begin{multline} 
\Im
D_{\mathrm{RPA}}=-\frac{N D_0^2\Im\Pi^R}{\fkt{}
{1+ND_0\Re\Pi^R}^2+\fkt{}{ND_0\Im\Pi^R}^2}\\ \label{EqImRPAInt}
\end{multline}
[see also Eq. \eqref{EqDefImDRPA} in Appendix \ref{AppRates}].
In Fig. \ref{FigIDRPA}b we see the plasmon peak in the RPA interaction propagator,
which is strongly asymmetric and is suppressed around the light cone, so that $\Im D_{\mathrm{RPA}}=0$
exactly at $\Omega=Q$.

Let us now discuss the status of the RPA in graphene. Above we have introduced
the large number of independent flavors $N\gg 1$. At zero temperature, this is the only
parameter which justifies the RPA summation in the problem of interacting Dirac fermions at $\mu=0$.
Indeed, in view of the absence of the screening, the non-RPA diagrams are parametrically the same as those
included in the RPA series for $N\sim 1$.
Furthermore, as discussed in the
end of Sec. \ref{SecPolopzT}, the renormalization-induced curvature of the spectrum (which is also beyond the RPA)
may dramatically affect the results obtained within the RPA.

However at finite $T$, the $T$-induced screening of the interaction, Eq.~\eqref{EqScreeninglength}, restores
the validity of the RPA for $q\ll T/v_F$ ($Q\ll 1$) even for $N\sim 1$. Indeed,  the $1/q$-singularity of the long-range
Coulomb interaction is not compensated in the denominator of
Eq.~\eqref{EqImRPAInt}  because the polarization operator  at finite $T$ is no longer linear-in-$q$.
The situation
becomes similar to that in conventional metals with a finite Fermi-surface, where the RPA is justified
for $q\ll k_F$. In graphene the role of $k_F$ is played by $T/v_F$, which in effect establishes an analog of
a finite Fermi-surface. Therefore, for $q\ll T/v_F$ the RPA does sum up the most singular interaction-induced terms:
all other terms are non-singular because of the screening. This means that all the observables that are dominated
by the collisions with the momentum transfer smaller than $T/v_F$ can be calculated (even with the correct numerical prefactors)
within the RPA. The RPA result for those observables that are dominated by $qv_F\sim T$ is parametrically correct, but the value
of the prefactor can not be found using the RPA.

Below we employ the finite-$T$ RPA for calculation of various scattering rates in graphene. For the sake of generality, we keep $N$ as a parameter. In what follows we focus on the case $\alpha_g N\ll 1$, but
whenever the rate under the consideration is dominated by $qv_F\alt T$, the condition $N\gg 1$ can be removed
so that we are allowed to use the RPA for $N\sim 1$.

\section{Scattering rates}
\label{SecRates}
In this section we calculate various inelastic scattering rates in clean graphene at finite temperature.
More specifically, we focus on the quantum scattering rate (and dephasing) and
the energy relaxation rate induced by the RPA-screened Coulomb interaction. In Sec. \ref{SubsecTransport} below, we will also
calculate the transport scattering rate due to inelastic collisions.
The quantum scattering rate determines the lifetime of quantum states (plane waves) and is related to the dephasing rate. The energy relaxation rate governs the relaxation of the quasiparticle distribution function. Finally, the transport scattering rate describes  the influence of the inelastic scattering on transport phenomena.

Although the origin of all these rates is the same---the inelastic electron-electron
collisions, these rates,
may strongly differ from each other.
For instance, this is exactly what happens in diffusive metals \cite{AA,Blanter}
because of the infrared-singular collision kernel.
Another prominent example of a non-trivial behavior of relaxation rates
related to the infrared singularities is a Luttinger liquid (disordered or clean).\cite{GMP,Ayash,Bagrets,Gutman}
On the other hand, within the Fermi-liquid theory of clean metals all the inelastic scattering rates behave in the same way, since the characteristic frequency/momenta transfer in the course of electron-electron collisions is determined by temperature.
The goal of this section is to understand whether the situation in clean graphene is similar to the Fermi-liquid or not.

We have already mentioned in the introduction (Sec. \ref{SecInt}) the previous works on the inelastic quantum scattering rate in graphene~\cite{Gonzalez94,Sarma07}. These works addressed the scattering rate at the GR level at zero temperature and
obtained the Fermi-liquid-type result $\tau_{\mathrm{q}}^{-1}\sim \alpha_g^2 N \epsilon\ll \epsilon$.
Naively, one could think that at finite temperature
this consideration would lead to $\tau_{\mathrm{q}}^{-1}\sim \alpha_g^2 N T$. However, since the GR result is completely determined by the ``mass shell'' ($\omega=v_Fq$) (see discussion in Sec.\ref{SecPolopzT}) one concludes that the RPA-resummation (which kills the on-shell interaction) would yield $\tau_{\mathrm{q}}^{-1}(\epsilon)=0$ for $T=0$ and arbitrary $\epsilon$.
Therefore, the finite-$T$ expectation based on the GR is also doubtful.

How the quasiparticle broadening behaves in the experimentally relevant
case of finite temperature is thus by far not obvious. As discussed in Sec.\ref{SecPolopfT}, the finite density of states at non-zero $T$ leads to screening of the Coulomb interaction, thus justifying the use of the RPA  which then sums up the
most singular contributions of the interaction for small momenta, $qv_F<T$.
In this section we find the behavior of inelastic rates
at finite temperature within the RPA.
As we have already seen in section \ref{SecPolopfT}, the behavior of the polarization operator at finite temperature is highly non-trivial. This leads to a rather rich behavior of the scattering rates.

\begin{figure}
 \includegraphics[width=0.9\columnwidth]{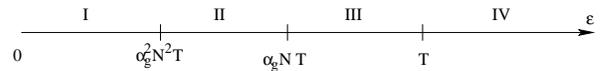}
\caption{Characteristic energy scales separating domains of distinct behavior of the rates.}
\label{FigScales}
\end{figure}

It turns out that, in addition to the temperature scale, two more characteristic
scales appear which are relevant for relaxation rates:
$\alpha_g^2N^2 T$ and $\alpha_gN T$.
In this section we will distinguish between the four regimes (I-IV) as  shown in Fig.~\ref{FigScales}.
The contributions to each of the scattering rates from different regions (1-4, see Fig. \ref{FigFinPolopAreas}) of the momentum-frequency plane are calculated separately.

\subsection{Quantum scattering rate in graphene}
\label{SecQscatt}
One of the main manifestations of the inelastic scattering
in electronic systems is the interaction-induced dephasing.
In order to analyze the effects of dephasing in graphene, we will follow the route
suggested by earlier works on 2D (in particular, diffusive)
systems.\cite{AA} A natural first step is to
calculate the quantum scattering rate $\tau_{q}^{-1}$ which is given by the imaginary part of
the quasiparticle self-energy taken at the mass-shell. Indeed, for
conventional metals with parabolic dispersion, in the high-temperature (ballistic) regime,
the dephasing rate $\tau_\phi^{-1}$ to leading order is given by
$\tau_{\mathrm{q}}^{-1}$, see Ref. \onlinecite{Narozhny}.
In the diffusive regime, the scattering kernel acquires an infrared
singularity leading  to a divergent $\tau_{\mathrm{q}}^{-1}$ at finite $T$.
This is a manifestation of the fact that the single-particle self-energy is not a gauge-invariant
object; no divergencies occur in observable (gauge-invariant) quantities, such as, e.g., dephasing rate.
However, even when the quantum scattering rate diverges,
the calculation of it turns out to be instructive: a
parametrically correct result for the dephasing rate can be obtained from the
expression for $\tau_{\mathrm{q}}^{-1}$ supplemented with an appropriate infrared
cutoff. It is thus useful to begin with analyzing $\tau_{\mathrm{q}}^{-1}$.

\subsubsection{Quantum scattering rate: definitions}
\label{SubSecQscatt}
The peculiarity of graphene is that the self-energy is a matrix
in the sublattice space, which has a non-zero imaginary part, $\Im\Sigma_v$ [Eq.~\eqref{ImSigmav}], 
in the off-diagonal components. 
Therefore, the definition of the quantum scattering time in graphene is actually not unique. 
Indeed, one can
associate with the quantum scattering rate the on-shell value of the imaginary correction to 
the energy in the full Green's function,
$\Im\Sigma_\epsilon$ [Eq.~\eqref{ImSigmaE}], similarly to the conventional Fermi-liquid theory:
\begin{multline}
 \frac{1}{2\tau_{\mathrm{q}}(\epsilon)}=-\theta(\epsilon)\left.\Im\Sigma_{\epsilon}(v_Fp,\vec{p})\right|_{p=\epsilon/v_F}
\\
-\theta(-\epsilon)\left.\Im\Sigma_{\epsilon}(-v_Fp,\vec{p})\right|_{p=-\epsilon/v_F}
\label{tauFLdef}
\end{multline}
Here at $\epsilon>0$ we have taken the self-energy at the ``$+$'' mass-shell corresponding 
to the positive energies, $\epsilon=v_F p$,
and at $\epsilon<0$ on the ``$-$'' mass-shell.
Clearly, 
\begin{equation}
  \frac{1}{\tau_{\mathrm{q}}(\epsilon)}=\frac{1}{\tau_{\mathrm{q}}(-\epsilon)}
\end{equation}
for undoped graphene ($\mu=0$) because of the particle-hole symmetry.
Within the RPA the explicit expression for the imaginary part of the total self-energy $\Sigma_{\epsilon}$ taken
at $\epsilon=v_F p$ reads:
\begin{multline}
\Im\Sigma_{\epsilon}(v_Fp,\vec{p})
= -\frac{\pi}{2} \int \frac{d\omega}{2\pi}\, \fktb{}{g(\omega)+f(v_F p-\omega)}\\ 
\times \int \frac{d^2q}{(2\pi)^2} \Im D^R(\omega,\vec{q}) \\
\times \fktb{}{\fkt{\delta}{v_F p-\omega-v_F
       |\vec{p}-\vec{q}|}+\fkt{\delta}{v_F p-\omega+v_F
       |\vec{p}-\vec{q}|}}. \\
\label{tauFLRPA}
\end{multline}

Alternatively, one can introduce the lifetime of the $+$ and $-$ chiral states through the corresponding self-energies, Eq.~\eqref{ImSigmaPM}:
\begin{equation}
 \frac{1}{2\tau_{\pm}(\epsilon)}=-\left.\Im\Sigma_{\pm}(\epsilon,\vec{p})\right|_{p=|\epsilon|/v_F}
\label{tauPMdef}
\end{equation}
which yields 
\begin{equation}
 \frac{1}{\tau_{+}(\epsilon)} = \frac{1}{\tau_{-}(-\epsilon)}.
\end{equation}
Using Eqs.~\eqref{ImSigmaPM} and \eqref{GlSpektralesgewicht}, 
we get for the self-energy of electrons ($+$ chirality)
\begin{multline}
\Im\Sigma_{+}(v_Fp,\vec{p})
= -\frac{\pi}{2} \int \frac{d\omega}{2\pi}\, \fktb{}{g(\omega)+f(v_F p-\omega)}\\ 
\times \int \frac{d^2q}{(2\pi)^2} \Im D^R(\omega,\vec{q}) \\
\times 
\left[\left(1+\frac{\vec{p}(\vec{p}-\vec{q})}{p|\vec{p}-\vec{q}|}\right)\fkt{\delta}{v_Fp-\omega-v_F
       |\vec{p}-\vec{q}|}\right.
\\
+\left. \left(1-\frac{\vec{p}(\vec{p}-\vec{q})}{p|\vec{p}-\vec{q}|}\right)\fkt{\delta}{v_Fp-\omega+v_F
       |\vec{p}-\vec{q}|}\right].\\
\label{tauPlusPlus}
\end{multline}
for the ``own'' mass-shell, and
\begin{multline}
\Im\Sigma_{+}(-v_Fp,\vec{p})
= -\frac{\pi}{2} \int \frac{d\omega}{2\pi}\, \fktb{}{g(\omega)+f(-v_F p-\omega)}\\ 
\times \int \frac{d^2q}{(2\pi)^2} \Im D^R(\omega,\vec{q}) \\
\times 
\left[\left(1-\frac{\vec{p}(\vec{p}-\vec{q})}{p|\vec{p}-\vec{q}|}\right)\fkt{\delta}{-v_Fp-\omega-v_F
       |\vec{p}-\vec{q}|}\right.
\\
+\left. \left(1+\frac{\vec{p}(\vec{p}-\vec{q})}{p|\vec{p}-\vec{q}|}\right)\fkt{\delta}{-v_Fp-\omega+v_F
       |\vec{p}-\vec{q}|}\right].\\
\label{tauPlusMinus}
\end{multline}
for the ``wrong'' (hole) mass-shell. 

The main formal difference between the two relaxation rates, Eq.~\eqref{tauFLdef} and Eq.~\eqref{tauPMdef}, which are related
by
\begin{equation}
 \frac{1}{\tau_{\mathrm{q}}(\epsilon)}=\frac{1}{2\tau_{+}(\epsilon)}+\frac{1}{2\tau_{-}(\epsilon)},
\label{meanrate}
\end{equation}
is in the appearance of Dirac factors $1\pm \cos\theta$ in Eqs.~\eqref{tauPlusPlus} and \eqref{tauPlusMinus}
where $\theta=\arccos(\vec{p}\vec{p}'/pp')$ is the scattering angle between the incoming momentum $\vec{p}$
and the momentum $\vec{p}'=\vec{p}-\vec{q}$ after scattering.
These factors are related to the additional Berry phase in the problem of Dirac particles, which arises due to the
overlap of Bloch functions and, in particular, forbids the backscattering within the same chirality and valley.
Note that for well-defined quasiparticles (i.e., in a Fermi-liquid situation) the self-energy at a ``wrong'' mass shell
would never be relevant. However, if the quasiparticle broadening is larger than the characteristic energy,
this is no longer the case, so that Eq.~(\ref{tauPlusMinus}) may then contribute to the observables.

In Eq.~\eqref{tauPlusPlus}, the term with $1+\cos\theta$ corresponds to the electron-electron scattering (right panel of Fig.~\ref{FigPolScetch}) which is determined by the contribution of region 3 in Fig. \ref{FigFinPolopAreas}. The
term with $1-\cos\theta$ is due to electron-hole scattering (electron-hole pairs, left panel of Fig.~\ref{FigPolScetch})
accompanied by the excitation of plasmons  
and is determined by the contribution of region 1 in Fig. \ref{FigFinPolopAreas}. The latter contribution is suppressed for the forward scattering $\theta=0$ because of the Dirac factor. Furthermore, at zero temperature only the electron-electron processes
are allowed by the kinematic restrictions. 

At finite $T$, however, the situation is different for the low-energy  domains
I and II ($\epsilon<\alpha_g N T$), where the contributions of the electron-electron and electron-hole scattering to the 
inelastic quantum scattering rates are of the same order. This means that the low-energy electron-type and hole-type
quasiparticles are strongly correlated by the mutual inelastic scattering, whereas at higher energies ($\epsilon>\alpha_g N T$,
corresponding to domains III and IV) the electronic and hole subsystems are only weakly coupled with each other, in agreement
with Ref.~\onlinecite{Foster09}.

As we will see below, depending on the energy range,
Fig. \ref{FigScales}, both $\tau_{\mathrm{q}}^{-1}$ and $\tau_{\pm}^{-1}$ may be larger or smaller than energy.
When the quasiparticle's broadening is small (``Fermi-liquid regime''), the two rates coincide since the inelastic scattering
is dominated by small scattering angles $\theta$ and the Dirac factors reduce to $0$ and $1$. At the lowest energies (domains I and II), both $\tau_{\mathrm{q}}^{-1}$ and $\tau_{\pm}^{-1}$ exceed the energy  which makes the notion of mass shell not well-defined. In particular, due to the inelastic broadening, the electronic excitation
(+ chirality) has tails at negative energies which overlap with the hole mass shell. 

Furthermore, as shown in Sec. \ref{SecErel}, the characteristic rate of the energy relaxation (energy mixing due to the diffusion over energy) is in the 
``non-Fermi-liquid regime'' of the same order as the quantum scattering rate, so that the electronic excitations 
constantly explore the hole mass shell and vice versa.
In this situation, the broadening of quasiparticles of a given chirality is in effect described by the total scattering rate $\tau_{\mathrm{q}}^{-1}$, according
to Eq.~\eqref{meanrate}, rather than Eq.~\eqref{tauPMdef}. Therefore, in what follows, we will mostly focus on the 
Fermi-liquid-type total rate $\tau_{\mathrm{q}}^{-1}$, formally defined in Eq.~\eqref{tauFLdef}.

\subsubsection{Quantum scattering rates: results and discussion}
\label{SubSecQscattRes}

\begin{figure}
 \includegraphics[width=\columnwidth]{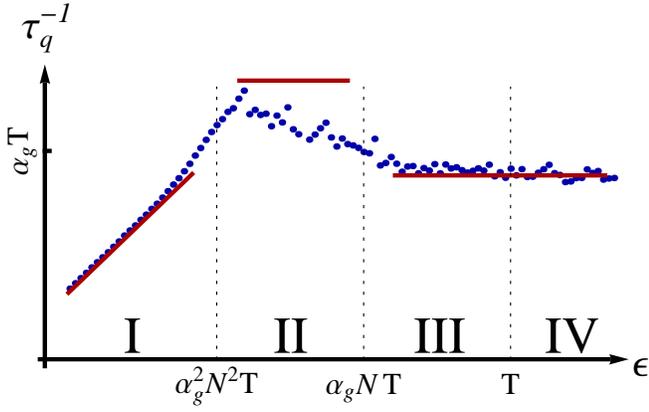}
\caption{(Color online) Quantum scattering rate for $\alpha_g N= 4\times 10^{-3}$ (double logarithmic scale).
Dots: exact values obtained by numerical evaluation; solid lines: analytical asymptotics, Eq.~\eqref{EqQScattRate}.
}
\label{FigTauq}
\end{figure}

The evaluation of integrals involved in the calculation of the total quantum scattering rate $\tau_{\mathrm{q}}^{-1}$ is outlined in Appendix \ref{AppRates}.
The result depends on the energy range (see Fig. \ref{FigScales}):
\begin{equation}
\frac{1}{2\fkt{\tau_{\mathrm{q}}}{\epsilon}}\approx \left\{\begin{array}{rrclc}
\displaystyle c_{1} \frac{T}{N}\sqrt{\frac{\epsilon}{T}}, \hspace*{0.2cm}& &\!\!\!\epsilon &\!\!\!\ll \alpha_g^2N^2T, &\!\!\mathrm{I} \\[0.5cm]
c_{2}\alpha_g T, \hspace*{0.2cm} & \alpha_g^2N^2T \ll &\!\!\!\epsilon &\!\!\ll \alpha_gN T,  &\!\!\!\mathrm{II} \\[0.5cm]
c_{3} \alpha_g T,\hspace*{0.2cm} &  \alpha_gNT \ll &\!\!\!\epsilon.& &\!\!\!\!\!\!\!\!\!\!\!\mathrm{III},\mathrm{IV}	
                        \end{array}\right.
\label{EqQScattRate}
		      \end{equation}
Here $c_{1}\sim 1$, $c_{2}=3\pi/2$, and $c_{3}=2\beta(2)/\pi$ are the
numerical coefficients of order unity, $\beta(x)$ is the Dirichlet beta function and $\fkt{\beta}{2}$ is the Catalan constant.
Note that for energies in Regimes II, III, and IV, the number of independent flavors $N$ drops out from the expression for $\tau_{\mathrm{q}}$.
The comparison of the asymptotic expressions \eqref{EqQScattRate} with the exact numerical
evaluations is shown in Fig. \ref{FigTauq}).

The obtained rates are dominated by different values of momenta transferred during the collision.
Furthermore, depending on the energy range the main contribution may come from different regions of the $Q$ vs. $\Omega$ plane
(Fig. \ref{FigFinPolopAreas}) as shown in Table \ref{TableQRateDetbyOrder}.

\begin{table}
\begin{center}
 \begin{tabular}{|c|cccc|}
 \hline
 $\tau_{\mathrm{q}}$ & I & II & III & IV \\
 \hline $qv_F\sim \omega$   &\hspace*{0.5cm} $T$\hspace*{0.4cm} &\hspace*{0.3cm} $(\alpha_gNT)^2/\epsilon$\hspace*{0.3cm} &\hspace*{0.31cm} $\alpha_gNT$\hspace*{0.3cm} &\hspace*{0.3cm} $\alpha_gNT$\hspace*{0.3cm} \\
 \hline
Regions  & 1,2,3,4 & 1,3 & 3 & 3 \\
\hline
 \end{tabular}
\caption{Momentum/frequency scales and regions of the $Q$-$\Omega$ plane that dominate the quantum scattering rate $\tau_{\mathrm{q}}^{-1}$
in different domains (I,II,III, and IV) of energy $\epsilon$.}\label{TableQRateDetbyOrder}
\end{center}
\end{table}

In Regime I the result is determined by momenta of order temperature and therefore all four regions contribute
in the same way: $\tau_{\mathrm{q}}^{-1}\sim (\epsilon T)^{1/2}/N$. All scattering angles $\theta$ contribute to
the result in Regime I. 
In order to evaluate the numerical prefactor $c_{\mathrm{I}}$ one needs the knowledge of the screened interaction in the
crossover around $qv_F\sim T$, which is beyond the analytic approximations for the polarization operator used above.
For energies in Regime II the dominant contributions come from regions 3 (electron-electron scattering) and 1 (electron-hole scattering); in both of them all scattering angles $\theta$ contribute.
In Regimes III and IV the main contribution to the quantum scattering rate stems from the region 3 (electron-electron scattering) and is dominated by the 
forward scattering ($\theta \alt \alpha_g N T/\epsilon\ll 1$).

An important feature of the quantum scattering rate is its non-monotonic energy dependence, see Eq.~\eqref{EqQScattRate}
and Fig. \ref{FigTauq}: with increasing energy
the quasiparticle broadening first grows in Regime I, has a maximum at $\epsilon\sim\alpha_g^2 N^2 T$,
then decreases in Regime II, and finally becomes energy-independent.
The maximum of $\tau_{\mathrm{q}}^{-1}$ occurs due to the resonant emission/absorption of plasmonic excitations.
We will see below that the non-monotonicity of the energy dependence is related to the peculiar properties of the dynamical screening in graphene and is a characteristic feature of all inelastic scattering rates in graphene.

\begin{figure*}
 \includegraphics[width=14cm]{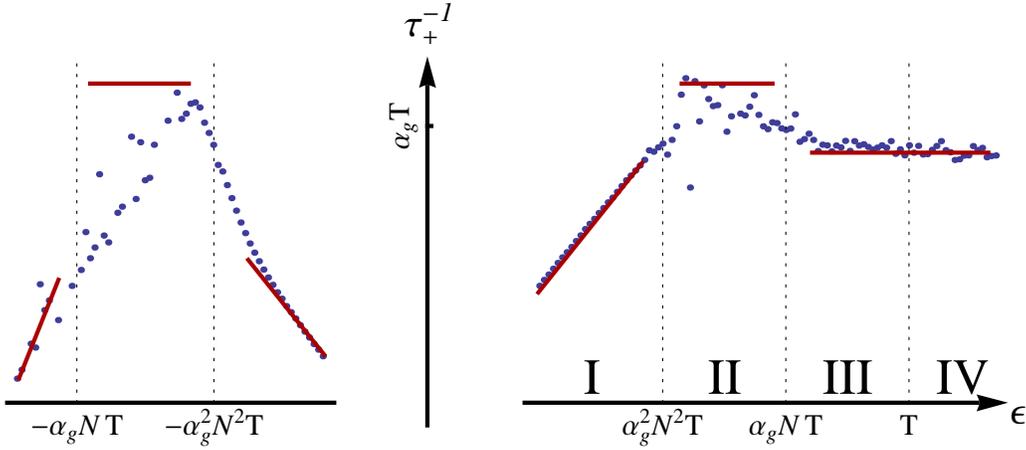}
\caption{(Color online) Quantum scattering rate $\tau_{+}^{-1}$ for the $+$ chirality (electrons) for $\alpha_g N= 4\times 10^{-3}$ (double logarithmic scale). Dots: exact values obtained by numerical evaluation; solid lines: analytical asymptotics 
(see Appendix \ref{AppRates}).
}
\label{FigTauQchiral}
\end{figure*}

In Fig. \ref{FigTauQchiral} we show the results for the quantum scattering rate of electrons, $\tau_+^{-1}$,
defined in Eq.~\eqref{tauPMdef}. One sees that Fig. \ref{FigTauQchiral} represents an ``unfolding'' of Fig. \ref{FigTauq}
into the contributions of ``own'' ($\epsilon>0$) and ``wrong'' ($\epsilon<0$) mass shells, according to Eq.~\eqref{meanrate}.
The hole mass shell with $\epsilon<0$ is probed by electrons due to the strong quasiparticle broadening.
At positive energies in Regimes III and IV ($\epsilon>\alpha_g N T$), the total scattering rate $\tau_{\mathrm{q}}^{-1}$ is 
dominated by the electronic scattering rate $\tau_{+}^{-1}$.

In contrast to the naive expectation ($\tau_{\mathrm{q}}^{-1}\sim \alpha_g^2 N T$) based on the Fermi GR result,
the quantum scattering rate calculated within the RPA is proportional to $\alpha_g$ (and does not depend on $N$) 
for $\epsilon\gg \alpha_g^2 N^2 T$ and is independent of $\alpha_g$ at the lowest energies. 
This enhancement of the inelastic scattering for $\alpha_g N\ll 1$
is a result of peculiar screening properties of graphene at finite $T$ which leads to a nonanalytic
behavior of the rate as a function of the natural four-fermion coupling constant $\alpha_g^2$. This behavior
bears a certain similarity to the behavior of the quantum scattering rate in a spinful Luttinger liquid.\cite{GMP,Ayash}

Importantly, the formal calculation of the total scattering rate $\tau_{\mathrm{q}}^{-1}$ in the lowest order
in the RPA interaction propagator yields $\tau_{\mathrm{q}}^{-1}(\epsilon)>|\epsilon|$ for regimes I and II
($\epsilon<\alpha_g N T$). This means that the quantum scattering rate is ill-defined in these low-energy domains.
This also signifies that for the calculation of observables in these energy domains the higher-order terms 
in the RPA screened interaction may become important. Nevertheless, the above lowest-order calculation in
domains I and II is useful as it provides the characteristic value for the intensity of inelastic processes.

Since the RPA quasiparticle broadening in domain I may overlap with
the higher-energy domains, where the quantum scattering rate is of the order of $\alpha_g T$, one can speculate that
the characteristic strength of inelastic scattering is given by $\tau_{\mathrm{q}}^{-1}(\epsilon)\sim \alpha_g T$ also at low
energies. Of course, the calculation based on the lowest-order RPA diagrams for the self-energy is then insufficient.
Moreover, the typical observables at finite temperatures are dominated by $\epsilon\sim T$ (the border between domains III 
and IV), where the quasiclassical broadening is smaller than energy and the above calculation is justified.
The low-energy inelastic relaxation may become relevant in the context of spectroscopy under strongly non-equilibrium conditions, 
for example, in problems related to tunneling into a non-equilibrium state (cf. Ref. \onlinecite{Gutman}).
In this situation the inelastic effects can be treated within the quantum-kinetic approach including non-RPA 
contributions, similarly to one-dimensional problems.\cite{Gutman}

\subsection{Dephasing rate at high temperatures}
\label{SecDeph}

Let us now discuss the dephasing rate. The result for the quantum scattering rate obtained for the clean graphene
allows us to evaluate the dephasing rate relevant to weak (anti)localization in the ballistic regime\cite{Tikhonenko09} of high temperatures $T\tau_{\mathrm{dis}}\gg 1$, where $\tau_{\mathrm{dis}}$ is the elastic mean free
time due to scattering off impurities. It is worth noting that this condition may coexist with the condition $\tau_\phi\gg\tau_{\mathrm{dis}}$
which allows long interfering paths in the weak-localization experiment. Indeed, the inelastic scattering is suppressed with
decreasing $\alpha_g$ (even though in a non-trivial way) so that one expects that $\tau_\phi$ can be made arbitrary long.

The result for the quantum scattering rate, Eq.~\eqref{EqQScattRate}, remains intact (up to the change in prefactor)
when we calculate it in a self consistent way as appropriate for estimating the dephasing rate:
\begin{equation}
 \tau_{\mathrm{\phi}}^{-1}\propto \nint{Q}{(T\tau_{\mathrm{\phi}})^{-1}}{\infty} \dots
\end{equation}
This happens due to the fact that the characteristic momenta dominating the integrals for the quantum scattering rate in Regimes II,III, and IV  is of the same order as the resulting rate [see Table \ref{TableQRateDetbyOrder} and Eq.~\eqref{EqQScattRate}], both are $\sim \alpha_g T$ for $N\sim 1$ (for $N\gg 1$, the characteristic momenta are much higher than $1/\tau_{\mathrm{q}}$). In Regime I of lowest energies, the characteristic momentum transfer is much higher than the rate. Therefore, in all these regimes the infrared
cutoff is redundant and the dephasing rate is given by the quantum scattering rate, Eq.\eqref{EqQScattRate},
similarly to the case of conventional metals in the ballistic regime~\cite{Narozhny} (although, in contrast to the conventional case,
the characteristic transferred frequencies are much smaller than temperature).
Since the characteristic energies involved in the transport coefficients are of order of $T$, we conclude that
in the ballistic regime the interference effects are governed by
\begin{equation}
 \frac{1}{\tau_\phi} \sim \alpha_g T.
\end{equation}
This prediction can be verified by transport experiments on graphene at sufficiently high temperatures (depending on the purity of the system, $T\agt 10-100 K$ for a typical setup with graphene deposited on a insulating substrate and $T\agt 1-10K$ for suspended graphene flakes). At lower temperatures $T\tau_{\mathrm{dis}}\ll 1$ corresponding to the ``diffusive regime'' with respect to interaction, one expect the conventional~\cite{AA,AleinerAltshulerGershenson} diffusive result for the dephasing rate
\begin{equation}
 \frac{1}{\tau_\phi} \sim \frac{T}{g}\ln g,
\end{equation}
where $g$ is the dimensionless Dirac-point conductance. At the Dirac point $g$ is close to unity and hence
$\tau_\phi^{-1}\sim T.$

\subsection{Energy relaxation rate}
\label{SecErel}

Let us now discuss the energy relaxation time in clean graphene.
In Sec. \ref{SecQscatt} we have seen that the typical momentum or energy transfer during the
electron-electron collision is much smaller than temperature.
In this situation, the energy relaxation occurs through multiple scattering processes which can be viewed
as diffusion in energy space (see, e.g., Ref.~\onlinecite{Levinson-book}).
The characteristic energy relaxation rate is given by the diffusion
coefficient of this problem, which amounts to introducing the factor $\mathcal{K_{\mathrm{E}}}=\omega^2/T^2$,
see Appendix \ref{defrates}, into the collision kernel.
More rigorous calculation of the energy relaxation or equilibritation rates can be done using the language of
kinetic equation; here we only estimate the typical rate $\tau_E^{-1}$ within the energy diffusion picture.

Once the quantum scattering rate is obtained, the calculation of the energy relaxation rate
can be done using the same steps as described in the previous section.
Technically, the integrals in Eq.~\eqref{tauFLRPA} are only slightly changed, which leads,
however, to a substantial difference between the two rates.
The detailed calculation can be found
in appendix \ref{AppRates}; here we present only the result:
	\begin{equation} \label{EqEnRelaxRate}
 		\frac{1}{\fkt{\tau_{E}}{\epsilon}}\sim\left\{ \begin{array}{crclc}
		\displaystyle		 \frac{T}{N} \sqrt{\frac{\epsilon}{T}},\
& \mathrm{I}\\[0.5cm]
\displaystyle	
		\!\frac{\alpha_g^2NT}{\sqrt{\epsilon/T}}\fkt{\log}{\!\frac{\epsilon/T}{\alpha_g^2N^2}\!},\
             & \mathrm{II},\mathrm{III}\\[0.5cm]
\displaystyle	
\alpha_g^2NT\fkt{}{\frac{\epsilon}{T}}^{3/2} \fkt{\log}{\frac{1}{\alpha_g N}},
& \mathrm{IV}
                     \end{array} \right.
	\end{equation}
Again the obtained rates are dominated by different momentum scales and by contributions of different regions as shown in table \ref{TableERateDetbyOrder}. Due to the factor $\mathcal{K_{\mathrm{E}}}$ all contributions except in regime IV are now determined by momenta of order temperature, which does not allow us to find the numerical value of the prefactors analytically.

Furthermore, in fact, the above calculation based on the energy diffusion is not justified for ``hot electrons'' with
high energies, $\epsilon>T$. Indeed, within our consideration, the characteristic energy transfer 
dominating the energy relaxation in Regime IV turns out to be much higher than $T$, in contrast to the original assumption.
Therefore, the estimate Eq.~\eqref{EqEnRelaxRate} for Regime IV can not be trusted and another approach is needed
for this Regime. 

\begin{table}[H]
\begin{center}
 \begin{tabular}{|c|cccc|}
 \hline
 $\tau_{\mathrm{E}}$ & I & II & III & IV \\
 \hline $qv_F\sim\omega $  &\hspace*{0.5cm} $T$\hspace*{0.5cm} &\hspace*{0.5cm} 
$T$\hspace*{0.5cm} &\hspace*{0.5cm} $T$\hspace*{0.5cm} &\hspace*{0.5cm} $\epsilon$\hspace*{0.5cm} \\
 \hline
Regions\hspace*{0.1cm}  & 1,2,3,4 & 1,2,3,4 & 1,2,3,4 & 4 \\
\hline
 \end{tabular}
\caption{Momentum/frequency scales and regions of the $Q$-$\Omega$ plane that dominate the energy relaxation rate $\tau_{\mathrm{E}}^{-1}$
in different domains (I,II,III, and IV) of energy $\epsilon$.}\label{TableERateDetbyOrder}
\end{center}
\end{table}
\begin{figure}
 \includegraphics[width=\columnwidth]{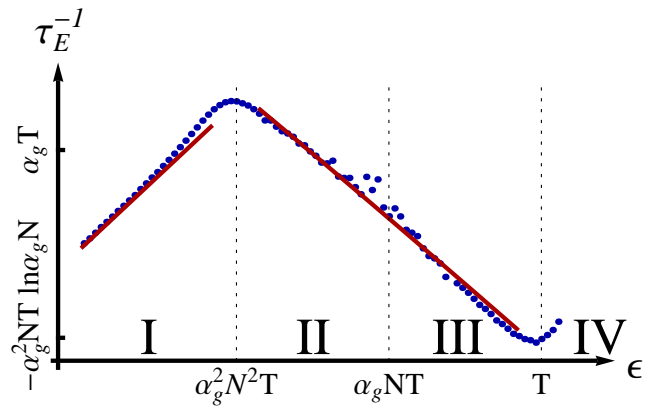}
\caption{(Color online) Energy relaxation rate for $\alpha_gN= 4\times 10^{-3}$ (double logarithmic scale), obtained from the energy-diffusion consideration.
Dots: exact values obtained by numerical evaluation; solid lines: analytical asymptotics, Eq.~\eqref{EqEnRelaxRate}.
Since the energy-diffusion approximation employed in the calculation is not justified in Regime IV, we do not present
the results in this Regime for $\epsilon\gg T$. }
\label{FigTe}
\end{figure}

The numerical results for the energy relaxation rate are shown in Fig.~\ref{FigTe} together with the analytical asymptotics, Eq.~\eqref{EqEnRelaxRate}. One sees that the energy relaxation has a minimum at $\epsilon\sim T$, where we recover, up to the logarithmic factor $|\ln\alpha_g|$, the GR result $\tau^{-1}_E\sim \alpha_g^2 N T$.
At lower energies the inelastic
scattering is enhanced due to the resonance in the RPA interaction propagator (the resonant condition correspond to $\epsilon\sim\alpha_g^2 N^2 T$), 
whereas at high energies the energy relaxation is stronger because of the large phase space available for inelastic processes.
We remind the reader, however, that at $\epsilon>T$ the above calculation based on the energy-diffusion approximation 
is not justified. In order to find
the correct relaxation of the distribution functions in this Regime of hot electrons, one should solve the
corresponding kinetic equation which can be reduced to the Fokker-Planck equation. This will be done elsewhere.

\section{Transport rate and conductivity}
\label{SecTrateanCond}
\subsection{Transport scattering rate due to inelastic collisions}
\label{SubsecTransport}
\begin{figure}
\begin{center}
 \includegraphics[width=\columnwidth]{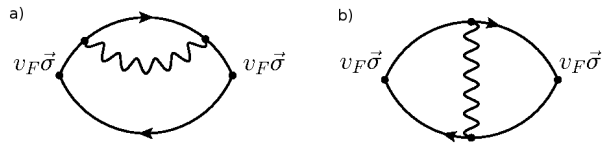}
\caption{Diagrams describing the first-order interaction correction to the conductivity}
\label{FigCondDiagsleading}
\end{center}
\end{figure}

In this section we calculate the transport relaxation rate due to inelastic collisions.
The expression for the corresponding kernel of the self-energy can be deduced from
the interaction-induced correction to the conductivity, as described below.
We are interested in the linear-response dc conductivity.
The leading order perturbative correction to the conductivity due to Coulomb interaction
is given by the two diagrams shown in Fig. \ref{FigCondDiagsleading}.

In the absence of interaction,  the conductivity of a clean graphene diverges at finite temperature.
Therefore, we have regularized the diagrams by introducing a small broadening ($\delta$) which mimics
the finite lifetime due to weak disorder. The general analytic expression of the lowest-order
interaction correction to the conductivity within the Keldysh formalism can
be found in Ref. \onlinecite{AleinerAltshulerGershenson}. 
This correction can be split into two parts (Fig. \ref{FigCondDiagsleading}a and Fig. \ref{FigCondDiagsleading}b,
respectively):  $\delta\sigma=\delta\sigma^s+\delta\sigma^v$, where
\begin{widetext}
\begin{multline}
\delta\sigma_{\beta\beta}^s=-\frac{1}{4i}\pintd{p_1}{2}{}{}\pint{\epsilon_1}{}{}
\pintd{p_2}{2}{}{}  \pint{\epsilon_2}{}{}\pd{f_1}{\epsilon}
\fktc{\mathrm{Tr}}{\vphantom{\int}j_{\beta}\fktb{}{\fkt{G^R}{1}-\fkt{G^A}{1}}j_{\beta}\right.\\
\left.\times \fktb{}{\vphantom{\int}\fkt{}{f_2+g}\fktb{}{D^R-D^A}
\fktb{}{\fkt{G^R}{1}\fkt{G^R}{2}\fkt{G^R}{1}-\fkt{G^A}{1}\fkt{G^A}{2}\fkt{G^A}{1}}\right.\right.\\
 +  \left.\left. f_2
\fktb{}{\fkt{G^R}{1}\fktb{}{\fkt{G^R}{2}D^R-\fkt{G^A}{2}D^A}\fkt{G^R}{1}-\fkt{G^A}{1}\fktb{}{\fkt{G^R}{2}D^R-\fkt{G^A}{2}D^A}
\fkt{G^A}{1}\vphantom{\sum}}\vphantom{\int}}}
\end{multline}
is the self-energy contribution and
\begin{multline}
\delta\sigma_{\beta\beta}^v= -\frac{1}{4i}
\pintd{p_1}{2}{}{}\pint{\epsilon_1}{}{}
\pintd{p_2}{2}{}{}
\pint{\epsilon_2}{}{}\pd{f_1}{\epsilon}
\fktc{\mathrm{Tr}}{\vphantom{\sum}j_{\beta}\fkt{}{f_2+g}\fktb{}{D^R-D^A} \right. \\
\times \left. \fktb{}{\fkt{G^R}{1}\fkt{G^R}{2}-\fkt{G^A}{1}\fkt{G^A}{2}}j_{\beta}\fktb{}{\fkt{G^R}{2}\fkt{G^R}{1}-\fkt{G^A}{2}\fkt{G^A}{1}}
\right. \\ + \left. 2j_{\beta}f_2
\fktb{}{\fkt{G^R}{1}-\fkt{G^A}{1}}\fktb{}{D^A\fkt{G^R}{2}j_{\beta}\fkt{G^R}{2}-D^R\fkt{G^A}{2}j_{\beta}\fkt{G^A}{2}}j_{\beta}
\fktb{}{\fkt{G^R}{1}-\fkt{G^A}{1}}\vphantom{\sum}}\vphantom{\int}
\end{multline}
is due to the vertex correction. Here $j_\beta=ev_F\sigma_\beta$ is the current operator in graphene and we used the short-hand notations for the arguments of Green's functions: $1=\vec{p_1},\epsilon_1$ and $2=\vec{p_2},\epsilon_2$. We first trace out the sublattice structure and then regularize the divergent integrals by $\delta$.
The most divergent interaction-induced correction to the dc conductivity is then
proportional to $\delta^{-2}$:
\begin{multline} \label{EqCondCorrwithReg}
  \delta\sigma=-\frac{ e^2 v_F^2}{4\delta^2}\pint{\epsilon_1}{}{}\pd{f}{\epsilon_1}\pint{\epsilon_2}{}{}
\pintd{p_1}{2}{}{}
\pintd{p_2}{2}{}{}
\,
\pi \fktb{}{\fkt{\delta}{\epsilon_1-v_F
       p_1}+\fkt{\delta}{\epsilon_1+v_F
       p_1}}
\mathrm{sign}\epsilon_1 \\ 
\times \underbrace{\fktb{}{1-\fkt{}{\frac{\vec{p}_1 \!\cdot\vec{p}_2}{p_1 p_2}}}}_{\text{Transport factor}} \underbrace{\fktb{}{1+\fkt{}{\frac{\vec{p}_1 \!\cdot\vec{p}_2}{p_1 p_2}}}}_{\text{Dirac factor}}\mathrm{K}_{\Sigma}(\epsilon_1,\epsilon_1-\epsilon_2,\vec{p_1},\vec{p_1}-\vec{p_2}).
\end{multline}
\end{widetext}
Here $\mathrm{K}_{\Sigma}$ is the integral kernel of the self-energy:
\begin{equation}
\fkt{\Im\,\Sigma^R_0}{\epsilon,\vec{p}}=\pintd{q}{2}{}{}\pint{\omega}{}{}
\fkt{\mathrm{K}_{\Sigma}}{\epsilon,\omega,\vec{p},\vec{q}} \label{EqDefKSigma}
\end{equation}

The correction, Eq.~\eqref{EqCondCorrwithReg}, is determined by the inelastic electron-electron scattering.
In the diagrams of the leading order in bare interaction shown in Fig. \ref{FigCondDiagsleading},
we have $\mathrm{K}_{\Sigma}\propto \Im D_0=0$, so that the inelastic corrections are zero.
What remains in the first-order conductivity correction, Fig. \ref{FigCondDiagsleading}, are the contributions responsible for the renormalization of the Fermi velocity in graphene coming from the real part of the self-energy;
note that these contributions are less singular in $\delta^{-1}$.

\begin{figure}
 \includegraphics[width=\columnwidth]{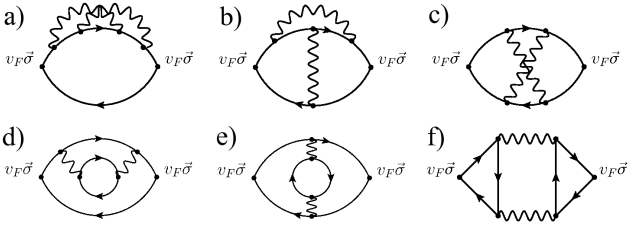}
\caption{Second-order skeleton diagrams for the conductivity}
\label{FigCondDiagssecond}
\end{figure}

\begin{figure}
 \includegraphics[width=\columnwidth]{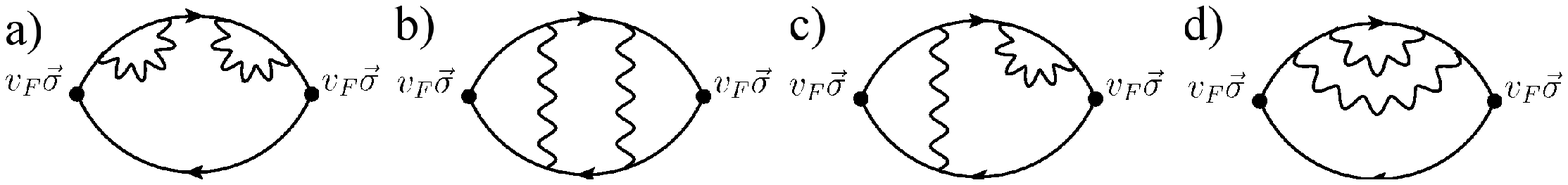}
\caption{Examples of second-order diagrams for the conductivity that contribute to the renormalization of velocity.}
\label{FigCondDiags2}
\end{figure}

Thus, we have to consider higher order corrections.
In Fig. \ref{FigCondDiagssecond} one can see all classes of second order skeleton diagrams.
There are also the second-order diagrams
of the ladder type, see Fig. \ref{FigCondDiags2}.
Such diagrams contribute only to the renormalization of $v_F$, similarly to the diagrams in Fig. \ref{FigCondDiagsleading},
and their contribution has been already included into the calculation simply by the replacement $v_F\to v_F(T)$.

In the large $N$ approximation diagrams in Fig. \ref{FigCondDiagssecond}d, \ref{FigCondDiagssecond}e, and \ref{FigCondDiagssecond}f dominate.
Diagram \ref{FigCondDiagssecond}f is known as Coulomb drag diagram and yields   a zero contribution
at the Dirac point due to electron-hole symmetry (see Refs. \onlinecite{Narozhny-drag,DasSarma-drag} and Appendix \ref{GR}).
Thus we are left with diagrams \ref{FigCondDiagssecond}d and \ref{FigCondDiagssecond}e which correspond to the two diagrams shown in
Fig. \ref{FigCondDiagsleading} with the second-order correction to the interaction instead of the bare one.
For these two diagrams Eq. \eqref{EqCondCorrwithReg} still holds with $\mathrm{K}_{\Sigma}\propto \Im D_1$.
This suggests to replace the bare interaction lines shown in Fig. \ref{FigCondDiagsleading} by the RPA interaction lines,
which would correspond to Eq. \eqref{EqCondCorrwithReg} with $\mathrm{K}_{\Sigma}\propto \Im D_{\mathrm{RPA}}$.
Note that even with the RPA-dressed interaction,  Eq. \eqref{EqCondCorrwithReg} still yields a divergent contribution
which we have regularized with $\delta$.

Considering the two independent scattering processes with the transport 
scattering rates $\delta$ and $\tau_{\mathrm{tr}}^{-1}$ such that
we find the Drude conductivity in the form
\begin{equation} \label{EqDrudeApprox}
 \sigma=e^2
\int{d\epsilon\,}  \fkt{\rho}{\epsilon} \left(-\pd{n_F}{\epsilon}\right)\,\frac{v_F^2\tau_{\mathrm{tr}}(\epsilon)}{2[1+\tau_{\mathrm{tr}}(\epsilon)\delta]},
\end{equation}
where $n_F(\epsilon)=[1-f(\epsilon)]/2$ is the thermal Fermi distribution function and $\rho(\epsilon)$ is the thermodynamic
density of states.
Expanding this formula in $(\delta\tau_{\mathrm{tr}})^{-1},$ we get the ``interaction-induced'' correction:
\begin{equation} \label{corrDrude}
 \delta\sigma=\frac{e^2v_F^2}{4 \delta^2}
\nint{\epsilon\,}{}{}\pd{f}{\epsilon}\,\frac{\fkt{\rho}{\epsilon}}{\tau_{\mathrm{tr}}(\epsilon)}.
\end{equation}

Comparing Eq.~\eqref{corrDrude} with Eq.~\eqref{EqCondCorrwithReg} we can identify
the interaction-induced transport scattering rate in graphene. 
The transport scattering rate obtained from the conductivity correction
corresponds to the kernel $\mathcal{K_{\mathrm{tr}}}$ defined in Appendix \ref{defrates}.

Furthermore, the connection between the kernels in the transport scattering rate and
the quantum scattering rate also follows from Eqs. \eqref{EqCondCorrwithReg}, \eqref{EqDefKSigma}, and \eqref{corrDrude}.
One can see from Eq. \eqref{EqCondCorrwithReg} that in the transport scattering rate
not only the contribution of forward scattering processes
is suppressed as in conventional systems but also the contribution of the
backward scattering. The latter suppression is due to the Berry phase of $\pi$ in graphene.

In the above derivation we have assumed $\delta\gg \tau_{\mathrm{tr}}^{-1}$ which allowed us to extract 
$\tau_{\mathrm{tr}}^{-1}$ from the expansion of the conductivity in $\tau_{\mathrm{tr}}^{-1}$. 
Using the generalized GR approach, we show in Appendix \ref{GR} that the expression 
for the transport scattering rate obtained in this way remains valid also for $\delta\to 0$.

\subsection{Transport scattering rate: results}

\begin{figure}
 \includegraphics[width=\columnwidth]{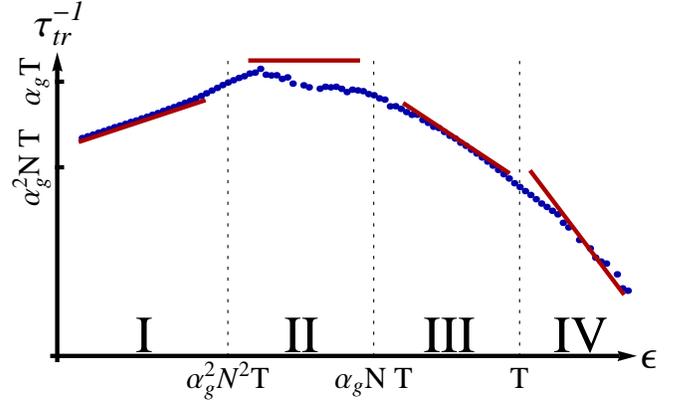}
\caption{(Color online) Transport scattering rate for $\alpha_g N=4\times 10^{-3}$ (double logarithmic scale).
Dots: exact values obtained by numerical evaluation; solid lines: analytical asymptotics, Eq.~\eqref{EqTransScatt}.}
\label{FigTtr}
\end{figure}

Similarly to the quantum scattering rate, the transport rate is dominated by region 3
in the regimes II, III, and IV.  Again, this is not so for energies that are in the regime $\mathrm{II}$.
In this case, we have an additional contribution from region 1, like in section \ref{SecQscatt}.
The calculation outlined in appendix \ref{AppRates} yields
\begin{eqnarray}\label{EqTransScatt}
 \frac{1}{\fkt{\tau_{\mathrm{tr}}}{\epsilon}}\hspace*{-0.2cm}
 &\sim& \hspace*{-0.2cm} \left\{ \begin{array}{crclr}
\displaystyle
 \frac{T}{N}\sqrt{\frac{\epsilon}{T}},  
						&\qquad \mathrm{I}\\[0.5cm]
\alpha_g T , 		&\qquad \mathrm{II}\\[0.5cm]
\displaystyle
\alpha_g^2NT\fkt{}{\frac{T}{\epsilon}}, 	&\qquad \mathrm{III}\\[0.5cm]
\displaystyle
\alpha_g^2NT\fkt{}{\frac{T}{\epsilon}}^2, 
						 &\qquad \mathrm{IV}
                   \end{array} \right.\nonumber\\
\end{eqnarray}

The characteristic momenta dominating the transport scattering rate as well as
the relevant regions in $Q$-$\Omega$ plane as shown in Table \ref{TableTRRateDetbyOrder}.
The transport scattering rate in Regimes $\mathrm{II}$ and $\mathrm{III}$ is dominated by momenta much smaller than temperature,
which allows us to find the numerical prefactors (given in Appendix B).
Regimes $\mathrm{I}$ and $\mathrm{IV}$ is determined by momenta of order temperature, which does not allow us to find the numerical coefficient in these regimes.
The analytical asymptotics are plotted alongside with the exact numerical result in Fig. \ref{FigTtr}.
\begin{table}[H]
\begin{center}
 \begin{tabular}{|c|cccc|}
 \hline
$\tau_{\mathrm{tr}}$ & I & II & III & IV \\
 \hline $qv_F\sim \omega$  &\hspace*{0.5cm} $T$\hspace*{0.5cm} &\hspace*{0.3cm} 
$\alpha_g^2N^2T/\epsilon$\hspace*{0.3cm} &\hspace*{0.3cm} $\epsilon$\hspace*{0.3cm} &\hspace*{0.3cm} $T$\hspace*{0.3cm} \\
 \hline
Regions  & 1,2,3,4 & 1,3 & 3 & 3 \\
\hline
 \end{tabular}
\caption{Momentum/frequency scales and regions of the $Q$-$\Omega$ plane that dominate the transport scattering rate $\tau_{\mathrm{tr}}^{-1}$
in different domains (I,II,III, and IV) of energy $\epsilon$.}\label{TableTRRateDetbyOrder}
\end{center}
\end{table}
Similarly to Sec. \ref{SecErel}, we see a strong enhancement of the transport scattering rate in region II.
When energy approaches temperature, the GR result
\begin{equation}
 \tau^{-1}_{\mathrm{tr}}(\epsilon\sim T) \sim \alpha_g^2 N T
\label{tautrtyp}
\end{equation}
is reproduced.

Comparing the transport scattering rate with the energy relaxation rate,
\begin{equation}
\label{relation-log}
\frac{\tau_{\mathrm{tr}}}{\tau_{\mathrm{E}}}\propto \ln \frac{1}{\alpha_g N},
\end{equation}
 we observe that in the limit of small $\alpha_g$ the relaxation of energy due to the
inelastic collisions occurs much faster than the velocity relaxation,
$\tau_{\mathrm{tr}}\gg \tau_{\mathrm{E}}$.
The difference between the two rates comes from the fact that the forward scattering
in graphene is strongly enhanced. The RPA screening suppresses the contribution of scattering angles smaller
than $\alpha_g$ thus regularizing the logarithmically divergent contribution to the energy relaxation rate.
On the other hand, the transport factor in $\mathcal{K}_{\mathrm{tr}}$, see Eq. \eqref{EqCondCorrwithReg},
kills the contribution of the forward scattering to the transport scattering rate much more efficiently than the RPA
screening (this is the reason why the GR result for $\tau_{\mathrm{tr}}$ is parametrically correct), so that
no logarithmic factor arises in $\tau_{\mathrm{tr}}$.

The relation Eq.~\eqref{relation-log} justifies
the hydrodynamic approach~\cite{Mueller08,Mueller091,Mueller092,Foster09}:
the distribution functions of electrons and holes equilibrate within each type of carriers
much faster than the direction of the velocity is changed. As a result, the distribution
functions effectively depend only on the
velocity direction.
This means that the interaction-induced transport scattering rate entering the
observables should be averaged over the temperature window:
\begin{equation}
\braket{\fkt{\tau_{\mathrm{tr}}^{-1}}{\epsilon}}=\frac{\displaystyle \nint{\epsilon}{0}{\infty}
\fkt{\ \rho}{\epsilon} \pd{f}{\epsilon} \frac{1}{\tau_{\mathrm{tr}}(\epsilon)}}{\displaystyle \nint{\epsilon}{0}{\infty}
\fkt{\ \rho}{\epsilon} \pd{f}{\epsilon}}.
\label{tautr-averaged}
\end{equation}
Evaluating these integrals numerically using the exact expression for the RPA propagators,
we find
\begin{equation}
 \braket{\fkt{\tau_{\mathrm{tr}}^{-1}}{\epsilon}}\simeq 1.18 \alpha_g^2 N T.
\label{tautrav}
\end{equation}
As we will see below, this averaged value of the transport scattering rate can be substituted
into the Drude expression for the conductivity, which yields the parametrically correct result
for the collision-dominated conductivity of clean graphene.

\subsection{Collision-limited conductivity} \label{SubSectCollisionLimitedC}
Above, when calculating the conductivity, we have assumed that the finite lifetime 
of quasiparticles is provided by some artificially introduced broadening $\delta$ which mimics
disorder. The introduction of the artificial lifetime allowed us to identify the 
contribution of the inelastic collisions
to the transport scattering rate by considering the perturbative-in-interaction 
contributions to the conductivity.
 Let us now discuss the conductivity of clean graphene, or, more precisely, 
the conductivity of graphene in the regime
when the inelastic collisions dominate over disorder scattering.

The conductivity of graphene at the Dirac point is a rather intricate
quantity. In the absence of interaction, transport in the ballistic
limit shows remarkable peculiarities in graphene.\cite{graphene-review} The interplay between
vanishing density of states and vanishing scattering rate leads to the
non-universal conductivity that depends on the measurement details.
In particular, finite size clean graphene sample of the
``short-and-wide'' geometry shows a behavior analogous to that of a
normal diffusive metal. The zero-$T$ conductivity of such a setup
 is $\sigma=4\times e^2/\pi h$. The same value of
conductivity was predicted for an infinite sample with a large but
finite electron lifetime. A different result, $\sigma = e^2/2
h$, was found in the undoped graphene at a large frequency. At
any non-zero value of the chemical potential, the conductivity of clean
graphene is infinite. At finite temperature, energies within the
temperature window contribute to the conductivity. This implies that
the Dirac-point conductivity becomes infinite in the noninteracting case
at any non-zero $T$.

As has been shown in Refs. \onlinecite{Kashuba08,Fritz08,Foster09}, the conductivity of clean undoped
graphene becomes finite due to the inelastic electron-electron collisions.
The estimate for the collision-limited conductivity can be obtained by substituting
the typical value of interaction-induced transport scattering time, Eq.~(\ref{tautrtyp}),
and the typical density of thermally populated states, $\rho(T)\sim N T/v_F^2$, into the Drude formula, which yields
\begin{equation}
\sigma= \frac{e^2}{h}\rho(T) v_F^2 \tau_{tr}(T)\sim \frac{e^2}{h}\frac{N T}{v_F^2}v_F^2\frac{1}{\alpha_g^2 N T}\sim
\frac{e^2}{h}\frac{1}{\alpha_g^2}.
\label{qualsigma}
\end{equation}
Note that the explicit dependence on $T$, $N$, and $v_F$ drops out from this formula; however, the temperature
dependence appears in Eq.~(\ref{qualsigma}) implicitly through the renormalization of $\alpha_g$.
A more rigorous calculation of the collision-limited conductivity requires the analysis
of the kinetic equation.\cite{Kashuba08,Fritz08} Importantly, the fast energy relaxation
discussed above not only simplifies such an analysis, but also reduces the kinetic
equation to the hydrodynamic model~\cite{Sachdev08,Mueller08,Mueller091,Mueller092,Foster09}.

The consideration of Sec. \ref{SubsecTransport}, which allowed us to find the transport scattering rate from the expression for the
conductivity, relied on the perturbative treatment of the interaction. This assumes the following hierarchy
of the energy scales:
\begin{equation}
\tau^{-1}_{\mathrm{tr}}\ll \delta \ll T.
\end{equation}
The first inequality implies that the broadening of the Green's functions is due to the artificial ``disorder'' rate
$\delta$, whereas the second inequality establishes the ballistic regime which allows us to neglect the dressing
of interaction by disorder.
Since the characteristic frequency transfer in the transport scattering rate is of order of temperature,
the resulting $\tau^{-1}_{\mathrm{tr}}$ does not depend on $\delta$ as long as $\delta \ll T$. 
Furthermore, in Appendix \ref{GR} we have calculated $\tau^{-1}_{\mathrm{tr}}$ from the generalized GR approach
for $\delta=0$ and reproduced the transport kernel $\mathcal{K}_{\mathrm{tr}}$ from Appendix \ref{AppRates}.
Therefore,
we expect that the Drude formula, yielding Eq.~\eqref{qualsigma}, is applicable also
for $\tau^{-1}_{\mathrm{tr}}\gg \delta$, i.e. in the collision-dominated transport regime. This expectation is supported by the results of Refs. \onlinecite{Kashuba08,Fritz08,Foster09}.   

 Assuming the validity of the Drude formula for $\delta\ll \tau^{-1}_{\mathrm{tr}}$, we evaluate
 the conductivity using the transport scattering rate given by Eq. \eqref{tautrav}:
 \begin{equation}\label{EqCondRes}
 \sigma=\frac{e^2}{h}\frac{N\pi}{\braket{\fkt{\tau_{\mathrm{tr}}^{-1}}{\epsilon}}}\nint{\epsilon}{0}{\infty}
\fkt{\rho}{\epsilon}\pd{f}{\epsilon}\approx
 \frac{e^2}{h}\frac{0.58}{\alpha_g^2}
\end{equation}
This result has the same form as found in Refs. \onlinecite{Kashuba08,Fritz08} within the kinetic equation approach.
The only difference is in the numerical value of the prefactor.
It is worth noting here that the kinetic approach of Refs. \onlinecite{Kashuba08,Fritz08} was based on the
GR calculation of the self-energies. As we have seen in Sec. \ref{SubsecTransport}, the RPA result
for the transport scattering rate at relevant energies $\epsilon\sim T$ [as well as the energy-averaged characteristic
rate, Eq.~(\ref{tautr-averaged})] has the same form as given by the GR. However, the proper inclusion of
the finite-$T$ changes the prefactor in the transport scattering rate. This explains the difference between the numerical
prefactors between our result, Eq.~\eqref{EqCondRes}, and the result of Refs. \onlinecite{Kashuba08,Fritz08}.
We thus see that it is important to include the full RPA-interaction into the kinetic-equation approach.
This will be done elsewhere.

Finally, let us discuss on the qualitative level the effect of disorder on the
Dirac-point conductivity in graphene. For simplicity we set $N=1$ below.
In the absence of interaction, effect of disorder on transport in the
ballistic regime is highly unconventional and strongly depends on
the type of randomness.\cite{Schuessler09,titov09,Schuesler10,ostrovsky10}
On the other hand, as we have seen above, the interaction effects
are crucially important for the transport already in a clean system.
It is thus important to explore transport in realistic graphene structures, with both the electron-electron
interaction and disorder taken into account.

\begin{figure}
 \includegraphics[width=0.85 \columnwidth]{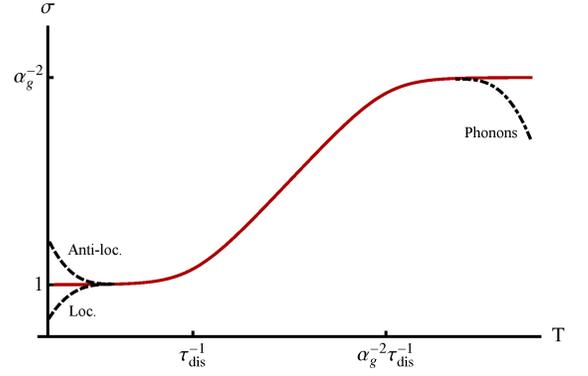}
\caption{Schematic plot (only parametrical scales are given) of the temperature dependence of the Drude conductivity (in units of $e^2/h$; solid line)
$\alpha_g\ll 1/\ln(\Lambda\tau_{\mathrm{dis}})$, where $\Lambda$ is the bandwidth.
For simplicity, the logarithmic temperature corrections to $\alpha_g$ which comes from the renormalization is not shown.
Dashed lines: low-$T$ behavior of the conductivity governed by quantum interference effects; depending on the character of disorder, the
localization, antilocalization, or criticality (coincides with the solid line) may occur. Dash-dotted line: expected high-$T$ behavior
governed by electron-phonon scattering.}
\label{Fig:Drude}
\end{figure}

\begin{figure}
 \includegraphics[width=0.9 \columnwidth]{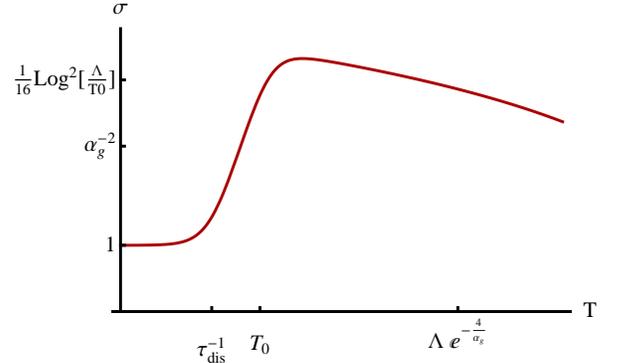}
\caption{Schematic plot of the temperature dependence of the Drude conductivity (in units of $e^2/h$)
for strong interaction, $\alpha_g\gg 1/\ln(\Lambda\tau_{\mathrm{dis}})$. The characteristic scale $T_0$ is given
by $T_0\sim \tau_{\mathrm{dis}}^{-1}\ln^2(\Lambda/T_0)\sim \tau_{\mathrm{dis}}^{-1}\ln^2(\Lambda\tau_{\mathrm{dis}})$.
For possible deviations at low and high $T$ (interference and phonon contributions, respectively), see Fig.~\ref{Fig:Drude}}
\label{Fig:strong}
\end{figure}

The role of disorder here
is twofold: (i) potential disorder introduces velocity relaxation, thus contributing to the
transport scattering rate:
\begin{equation}
\frac{1}{\tau_{tr}}\sim \alpha_g^2 T + \frac{1}{\tau_{\mathrm{dis}}},
\end{equation}
and (ii) establishes a finite density of states already in the Dirac point:
\begin{equation}
\rho\propto T+ \frac{1}{\tau_{\mathrm{dis}}}.
\end{equation}
Substituting these formulas into the Drude conductivity, we obtain the following result describing
the crossover between the collision-dominated and disorder-dominated regimes:
\begin{equation}
\sigma\sim \frac{e^2}{h}\frac{T+ \frac{1}{\tau_{\mathrm{dis}}}}{\alpha_g^2 T + \frac{1}{\tau_{\mathrm{dis}}}}.
\label{expect-sigma}
\end{equation}
This expected temperature dependence of the Drude conductivity is shown in Fig. \ref{Fig:Drude} for the case weak interaction
(or strong disorder),
$\alpha_g\ll 1/\ln(\Lambda\tau_{\mathrm{dis}})$ when the renormalization of $\alpha_g$ gives only small logarithmic corrections.
For stronger interaction (or weaker disorder), $\alpha_g\ll 1/\ln(\Lambda\tau_{\mathrm{dis}})$, the renormalization of
$$
\alpha_g(T)=\frac{\alpha_g}{1+\frac{\alpha_g}{4} \ln\frac{\Lambda}{T}},
$$
becomes strong [so that the renormalized coupling ``forgets'' about its bare value, $\alpha_g(T)\to 4/\ln(\Lambda/T)$] already in the collision-dominated regime,
see Fig.~\ref{Fig:strong}.
One sees that, since the interaction-induced transport rate contains $\alpha_g^2$ whereas
the density of states of thermally excited quasiparticles does not, the two crossover $T$-scales
appear, which establishes an intermediate regime of the ballistic transport,
$$\frac{1}{\tau_{\mathrm{dis}}}\ll T\ll \frac{1}{\alpha_g^2\tau_{\mathrm{dis}}}.$$
Again, a more rigorous derivation of the conductivity in the presence of both disorder
and inelastic scattering is based on kinetic-equation approach and will be performed elsewhere.
Note that we neglected the phonon contribution to the relaxation rates (for estimate of
their contribution, see, e.g. Ref. \onlinecite{Tikhonenko09}) which becomes relevant at sufficiently high temperatures.

At $T\sim 1/\tau_{\mathrm{dis}}$ the Drude conductivity becomes of the order of conductance quantum and
the dephasing rate becomes of the order of $T$.
At lower temperatures, the $T$ dependence of the conductivity
is governed by interference effects: localization, antilocalization, or critical behavior may occur, depending on the
symmetry of disorder.\cite{ostrovsky06,OurPapers,Anderson50} The crossover scale $1/\tau_{\mathrm{dis}}$ in typical experiments on high-quality graphene is in the range $T\sim 1-100 K$.

\section{Conclusions}
\label{SecConcl}

In conclusion, we have analyzed the inelastic electron-electron scattering in graphene
using the Keldysh diagrammatic approach.  We have demonstrated that finite
   temperature strongly affects the screening properties of graphene.
   This, in turn, dramatically influences the inelastic scattering rates as
   compared to the zero-temperature case.
We have calculated the finite-$T$
   quantum scattering rate, see Eq.~\eqref{EqQScattRate} and Fig.~\ref{FigTauq}, which
is relevant for dephasing of interference processes. We have identified an
hierarchy of regimes, Eq.~\eqref{FigScales}, arising due to the interplay of a plasmon enhancement of the scattering
and finite-temperature screening of the interaction.  The lifetime of quasiparticles with
energies close to the Dirac point has been found to be independent of the
coupling constant. We have further calculated the energy relaxation rate, Eq.~\eqref{EqEnRelaxRate} and Fig.~\ref{FigTe},
and transport scattering rate, Eq.~\eqref{EqTransScatt} and Fig.~\ref{FigTtr}.
For all the three rates, we have found a non-monotonic energy dependence which has been attributed to the resonant excitation of plasmons. Finally, we have discussed the collision-limited conductivity of clean graphene as well as the expected behavior
of the high-temperature conductivity in the presence of disorder, see Eqs.~\eqref{EqCondRes} and \eqref{expect-sigma}, respectively.
Our results complement the kinetic-equation and hydrodynamic
approaches for the collision-limited conductivity.

Our approach that employs the Keldysh formalism can be generalized
for the treatment of physics of inelastic processes
in strongly non-equilibrium setups. In particular, this framework is expected to allow us to investigate interaction effects on full counting  statistics of the electron transport in
graphene and to develop the theory of tunneling spectroscopy in
  strongly biased graphene setup.

\begin{acknowledgments}
We thank S. Carr, L. Fritz,  M. M\"uller, S. Ngo Dinh, and M. Titov for interesting and useful discussions.
The work was supported by the Center for Functional Nanostructures of the Deutsche
Forschungsgemeinschaft (DFG), by SPP ``Graphene'' of the DFG, by the Rosnauka grant 02.740.11.5072, and by the EUROHORCS/ESF
EURYI Award scheme (I.V.G.).

\end{acknowledgments}

\onecolumngrid

\begin{appendix}
\section{Polarization Operator}
\label{AppPolarisation-Operator}
\begin{figure}
\begin{center}
 \includegraphics[width=8cm]{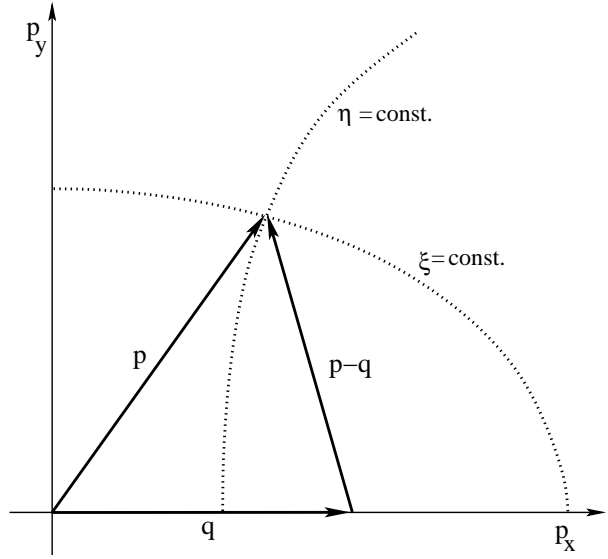}
\caption{Sketch of elliptic coordinates}
\label{FigElipticSketch}
\end{center}\end{figure}

In this Appendix we evaluate the polarization operator, Eq. (\ref{EqPolopstartCalc}), at finite temperature.
In a scattering process with emitting a photon, three different momenta are involved which form
in two dimensions triangles as shown in figure \ref{FigElipticSketch}.
Specifically, if an electron before scattering has the momentum $\vec{p}$ and the
emitted photon carries momentum $\vec{q}$,
the electron that is left over has to carry $\vec{p}-\vec{q}$.
The angular integration over the transferred momentum $\vec{q}$ becomes complicated.
To proceed further it is convenient to choose elliptic coordinates defined
by $\xi=p+\ABS{\vec{p}-\vec{q}}$ and $\eta=p-\ABS{\vec{p}-\vec{q}}$.
The corresponding coordinate system is shown in Fig. \ref{FigElipticSketch}.
Using the elliptic coordinates, the expressions for the imaginary and real parts of the polarization bubble
take the form
\begin{equation} \label{GlPolopSimplifyedIm}
 \Im\Pi^R  = \frac{\sinh(Q \beta)}{2 \pi}
\Re \left[ {\frac{Q}{\sqrt{\beta^2-1}}}\nint{\eta}{0}{1}
\frac{\sqrt{1-\eta^2}}{\cosh(Q \beta)+\cosh(\eta Q)} + {\frac{Q}{\sqrt{1-\beta^2}}}\nint{\xi}{1}{\infty}
\frac{\sqrt{\xi^2-1}}
{\cosh(Q \beta)+\cosh(\xi Q)}\right],
\end{equation}
\begin{equation} \label{GlPolopSimplifyedRe}
\Re\Pi^R = -\frac{Q}{\pi^2} 
\nint{\xi}{1}{\infty}
\nint{\eta}{0}{1} \,
\frac{\displaystyle\fkt{\sinh}{\xi Q}\sqrt{\frac{1-\eta^2}{\xi^2-1}}
\frac{\xi}{\beta^2-\xi^2}
+
\fkt{\sinh}{\eta Q}\!\sqrt{\frac{\xi^2-1}{1-\eta^2}}
\frac{\eta}{\beta^2-\eta^2}
}{\fkt{\cosh}{\xi Q}+\fkt{\cosh}{\eta Q}} .
\end{equation}
Here and below we introduce the notation $\beta=\Omega/Q$ to decouple expansions in small or large $Q$ from the behavior at the singularity $Q=\Omega$.
For simplicity we set $v_F=1$ and $T=1$.
The integrals in Eqs. (\ref{GlPolopSimplifyedIm}) and (\ref{GlPolopSimplifyedRe}) are evaluated separately in the four
regions shown in Fig. \ref{FigFinPolopAreas}.

\subsubsection{Region 1}\label{SectReg1}
In region 1 the condition $Q \ll 1$ and $\beta>1$ hold, which means that in Eq. \eqref{GlPolopSimplifyedIm}
only the part with the $\eta$-integral is left.
Expanding the integrands in small $\eta Q\ll 1$, we get
\begin{eqnarray} \label{EqImPolReg1}
 \Im \Pi^R  &\simeq& \frac{Q\sinh(Q \beta)}{2 \pi\sqrt{\beta^2-1}}\nint{\eta}{0}{1}\frac{\sqrt{1-\eta^2}}{\cosh(Q \beta)+1} =\frac{Q}{8 }\frac{\fkt{\tanh}{Q\beta/2}}{\sqrt{\beta^2-1}}
\end{eqnarray}
for imaginary part and
\begin{eqnarray} \label{EqRePolReg1}
\Re\Pi^R &\simeq&\frac{-Q}{\pi^2 } \overbrace{\int_{0\!\!}^{1\!\!}\!\!\frac{d\eta}{\sqrt{1\!-\!\eta^2}}\frac{Q\eta^2}{\beta^2\!-\!\eta^2}}^{=\frac{\pi Q}{2} \fkt{}{\frac{\beta}{\sqrt{\beta^2-1}}-1}}\overbrace{\nint{\xi}{1\!}{\infty\!\!\!\!\!}\frac{\sqrt{\xi^2-1}}{\fkt{\cosh}{\xi Q}\!+\!1}}^{\simeq\frac{2\ln 2}{Q^2}}\simeq -\frac{\ln 2}{\pi}\! \fkt{}{\frac{\ABS{\beta}}{\sqrt{\beta^2-1}}-1}
\end{eqnarray}
for the real part of $\Pi^R$.
The  term $\propto \fkt{\sinh}{\xi Q}$ in Eq. \eqref{GlPolopSimplifyedRe} yields a contribution of the order of $\sim Q^2$.
The second term $\propto \fkt{\sinh}{\eta Q}$ results in Eq. \eqref{EqRePolReg1}.

\subsubsection{Region 2}\label{SectReg2}
In region 2, we have $Q \gg 1$ and $\beta>1$. In Eq. \eqref{GlPolopSimplifyedIm} again only the $\eta$-integral is left and by neglecting $\fkt{\cosh}{\eta Q}$  and  expanding in large $Q$ we get
\begin{eqnarray}\label{EqImPolReg2}
 \Im  \Pi^R  &\simeq& \frac{1}{2 \pi}
\frac{\fkt{\sinh}{Q\beta}}{\sqrt{\beta^2-1}}\nint{\eta}{0}{1}\frac{\sqrt{1-\eta^2}}{\cosh(Q \beta)}= 
\frac{1}{8}\frac{Q \mathrm{sign} \beta}{\sqrt{\beta^2-1}}.
\end{eqnarray}
By expanding Eq.~\eqref{GlPolopSimplifyedRe} in large $Q$ and resolving exponentials in the denominator by a geometric series, we see that the leading contribution
to the real part of the polarization operator is small in region 2, being of the order of $\sim Q^{-2}\ll 1$.

\subsubsection{Region 3}\label{SectReg3}
Region 3 is characterized by the conditions $Q \ll 1$ and $\beta<1$. In this region, only the $\xi$-integral contributes
to Eq.~\eqref{GlPolopSimplifyedIm}.
By expanding the integrand in small $Q\beta$ we get
\begin{eqnarray} \label{EqImPolReg3}
 \Im  \Pi^R  &\simeq&
\frac{Q \beta }{2 \pi}
\frac{Q}{\sqrt{1-\beta^2}}
\overbrace{\nint{\xi}{1\!}{\infty\!\!\!\!}\frac{\sqrt{\xi^2-1}}{1+\!\cosh(\xi Q)}}^{\simeq\frac{2\ln 2}{Q^2}}
\simeq\frac{\ln 2}{\pi}\Re{\frac{\beta}{\sqrt{1-\beta^2}}}.
\end{eqnarray}
The simplification of Eq. \eqref{GlPolopSimplifyedRe} in region 3 is similar to that leading to Eq. \eqref{EqRePolReg1}.
The contribution of the first term in Eq. \eqref{GlPolopSimplifyedRe} is of order $Q\ln Q$, yielding
\begin{eqnarray} \label{EqRePolReg3}
\Re\Pi^R &\simeq & -\frac{Q}{\pi^2 }\!
\overbrace{\int_{0}^{1}\frac{d\eta}{\sqrt{1-\eta^2}}\frac{Q\eta^2}{\beta^2-\eta^2}}^{=-\frac{\pi Q}{2}}\, \overbrace{\nint{\xi}{1\!}{\infty}\frac{\sqrt{\xi^2-1}}{\fkt{\cosh}{\xi Q}+1}}^{\simeq \frac{2\ln 2}{Q^2}}\simeq \frac{\ln 2}{\pi}.
\end{eqnarray}

\subsubsection{Region 4}\label{SectReg4}
Finally, in region 4 the conditions $Q \gg 1$ and $\beta<1$ are fullfiled. In Eq. \eqref{GlPolopSimplifyedIm} only the $\xi$-integral contributes and by neglecting $\fkt{\cosh}{Q\beta}$  and expanding the integrand in large $Q\xi$ we get
\begin{eqnarray} \label{EqImPolReg4}
 \Im  \Pi^R  &\simeq &
\frac{\sinh(Q \beta)}{\pi}
	\frac{Q}{\sqrt{1-\beta^2}}
\underbrace{\nint{\xi}{1}{\infty}\sqrt{\xi^2-1}\,e^{-Q\xi}}_{\simeq e^{-Q}\frac{1}{Q^{3/2}}\sqrt{\frac{\pi}{2}}}
\simeq \frac{\fkt{\sinh}{Q\beta}}{\sqrt{\beta^2-1}}\frac{e^{-Q}}{\sqrt{2\pi \, Q}}.
\end{eqnarray}
In region 4 the simplification of Eq. \eqref{GlPolopSimplifyedRe} differs from that in region 2 in one important point.
By neglecting $\fkt{\cosh}{Q\eta}$ in comparison to $\fkt{\cosh}{Q\xi}$, the first term [$\propto \fkt{\sinh}{Q\xi}$] yields no principle value integral, while $\beta$ is smaller $1$. This results in
\begin{eqnarray}\label{EqRePolReg4}
\Re\Pi^R &\simeq & -\frac{Q}{\pi^2 }  \nint{\xi}{1}{\infty}\frac{\xi}{\sqrt{\xi^2-1}\fkt{}{\beta^2-\xi^2}}\nint{\eta}{0}{1}\sqrt{1-\eta^2}
=\frac{1}{8}\frac{Q}{\sqrt{1-\beta^2}}.
\end{eqnarray}

\section{Calculating the rates}\label{AppRates}
In this Appendix we calculate the integrals for inelastic scattering rates.

\subsection{Definitions of the rates}
\label{defrates}

The rates we are interested in are defined
by inserting the kernel $\mathcal{K_{\mathrm{j}}}$ into the integrand for
the imaginary part of the total self-energy, Eq.~\eqref{tauFLRPA}
($\mathrm{j}=\mathrm{q},+,\mathrm{E},\mathrm{tr}$):
\begin{itemize}
  \item Total quantum scattering-rate ($\mathrm{j}=\mathrm{q}$): 
$$\mathcal{K_{\mathrm{q}}}=1;$$
  \item Energy relaxation-rate ($\mathrm{j}=\mathrm{E}$): 
$$\mathcal{K_{\mathrm{E}}}=\frac{\omega^2}{T^2};$$
  \item Transport scattering-rate ($\mathrm{j}=\mathrm{tr}$): 
$$\mathcal{K_{\mathrm{tr}}}
=\frac{1}{2} 
\sin^2\theta=\frac{1}{2}\frac{q^2\sin^2\gamma}{\epsilon^2+q^2-2\epsilon 
q \cos\gamma }.$$
\end{itemize}
Here $\theta$ is the angle between incoming particle and outgoing 
particle and $\gamma$ is the angle between $\vec{q}$ and $\vec{p}$.
The origin of $\mathcal{K_{\mathrm{tr}}}$ is explained in Sec. 
\ref{SubsecTransport}.
The kernel for the chiral scattering rate $\tau_{\pm}^{-1}$ depends on the scattering channel.
In particular, for $+$ chirality at $\epsilon>0$ the electron-electron scattering kernel contains 
$\cos^2(\theta/2)$ and the electron-hole scattering kernel contains $\sin^2(\theta/2)$ due to 
Dirac factors.

It is convenient to introduce the dimensionless energy $y=\epsilon/2T$. 
The integrals we have to handle are of the following form:
\begin{equation} \label{EqScateringrateForm}
  \fkt{\tau^{-1}_{\mathrm{j}}}{y}
=-\frac{2 T^2}{v_F^2}\pint{Q}{0}{\infty}Q\pint{\gamma}{0}{2\pi}
\sum\limits_{\Omega=y\pm\sqrt{y^2+Q^2+2Qy\cos\gamma}}
\hspace*{-1cm}\fktb{}{\fkt{\mathcal{K_{\mathrm{j}}}}{\Omega,Q,y}\fkt{\Im 
D_{\mathrm{RPA}}^R}{\Omega,Q}\fktb{}{\fkt{\coth}{\Omega}+\fkt{\tanh}{y-\Omega}}}.
\end{equation}
Since the combinations $y\pm\sqrt{y^2+Q^2+2Qy\cos\gamma}$ lead to 
complicated integrands, we split the integrals into the parts 
corresponding to $Q\ll y$ and $y\ll Q$. Here $\gamma$ is the angle 
between the transferred momentum $\vec{q}$ and the initial
momentum $\vec{p}$.

\subsection{Simplifying the integrand}
\label{SectSimplIntegrands}
The imaginary part of the interaction propagator is given by
\begin{eqnarray} \label{EqDefImDRPA}
\fkt{\Im D_{\mathrm{RPA}}^R}{\Omega,Q}
=-\frac{\fkt{D_0^2}{Q}N\fkt{\Im\Pi}{\Omega,Q}}
{\fktb{}{1
+\fkt{D_0}{Q}N\fkt{\Re\Pi}{\Omega,Q}}^2+\fktb{}{\fkt{D_0}{Q}N\fkt{\Im\Pi}{\Omega,Q}}^2},
\end{eqnarray}
where
\begin{equation}
  \fkt{D_0}{Q}=\frac{v_F^2}{T}\frac{\alpha_g\pi}{Q}
\end{equation}
is the bare Coulomb interaction.

The splitting of the $Q$ integral leads to the following simplification:
\begin{eqnarray} \label{EqDefArtificialSimpl}
  \Omega\simeq y\pm\left\{\begin{array}{lr}y-Q\cos\gamma& Q\ll y\\ 
Q-y\cos\gamma& y\ll Q\end{array}\right.,&& 
\sqrt{\ABS{Q^2-\Omega^2}}\simeq\left\{\begin{array}{lr}\begin{array}{lc} 
2y &\qquad + \\ Q\ABS{\sin\gamma}\hspace*{0.8cm} &\qquad - \end{array} & 
Q\ll y \\ \begin{array}{lc} 2\sqrt{yQ}\ABS{\sin\frac{\gamma}{2}} &\qquad 
+ \\ 2\sqrt{yQ}\ABS{\cos\frac{\gamma}{2}} &\qquad - \end{array} & y\ll Q
  \end{array}\right.
\end{eqnarray}
Here $+,-$ correspond to the two possible values for $\Omega$ that 
appear in Eq.~\eqref{EqScateringrateForm}.
These signs reflect the separation of integration domains into the parts 
above and below
the mass-shell line $Q=\Omega$: $+$ corresponds to $\Omega>Q$ and $-$ 
corresponds to $\Omega<Q$.
Below we simplify the integrands 
$\mathcal{I}=\fktb{}{g+f}\mathcal{K_{\mathrm{j}}}\Im D_{\mathrm{RPA}}^R$ 
separately in each to the region in the $Q$ vs $\Omega$ plane.

\subsubsection{Region 1}
In region 1, the dominant contribution to all the rates comes from the 
domain $y\ll Q$. Using the simplified polarization operator, Eq. 
\eqref{EqSimpPolopatnonZT}, we obtain
\begin{equation}
  \fkt{\mathcal{I}_{1,2}^{\mathrm{j}}}{y,Q,\gamma}
\simeq \frac{v_F^2}{T}\frac{1}{N}\frac{\pi^2}{8}
\frac{\sqrt{yQ}\ABS{\sin\frac{\gamma}{2}}}
{\displaystyle \fkt{}{\frac{2\sqrt{y}\sqrt{Q}}{\alpha_gN}
\ABS{\sin\frac{\gamma}{2}}-\ln 2}^2
+\fkt{}{\frac{\pi}{16}Q^2}^2} \times
\left\{\begin{array}{ll} 1,& \mathrm{j}
=\mathrm{q} \\ 2\cos^2\frac{\gamma}{2},  &\mathrm{j}= +  \\ 4Q^2,  &\mathrm{j}=\mathrm{E} \\ \frac{1}{2} \sin^2\gamma, 
& \mathrm{j}=\mathrm{tr}\end{array}\right. .
\end{equation}
Here and below the first digit in the subscript of the function 
$\mathcal{I}$ denotes the region in the $Q$ vs $\Omega$ plane, while the 
second digit is $1$ for $Q\ll y$ and $2$ for $Q\gg y$.
\subsubsection{Region 2}
\label{Region2}
The contribution to all the rates coming from the region 2 is
exponentially small in Regime IV. In other Regimes, it contains at least 
an extra $\alpha_g$ as compared to the contributions
of region 3, except for the situations when results for the rate are 
determined by momenta $q$ of order $T/v_F$. In this situations, it turns 
out that the asymptotics produced by region 2 is the same as the 
asymptotics of region 3. We remind the reader that finding the numerical 
value of the prefactor is beyond our analytical approach when integrals 
are dominated by $q\sim T/v_F.$
Therefore, there is no case where we need to calculate the contribution 
of region 2.
\subsubsection{Region 3}
Region 3 appears to be the most important region because most of the 
final results for the rates are determined by this region.
In this region both small and large $y$ compared to $Q$ are important:
\begin{itemize}
  \item $y\gg Q$ :
\begin{equation}
\fkt{\mathcal{I}_{3,1}^{\mathrm{j}}}{y,Q,\gamma}\simeq\frac{v_F^2}{N T} 
\frac{\pi}{\ln 2}\frac{\ABS{\sin\gamma}Q^{-1}}{\displaystyle
\fktb{}{\fkt{}{\frac{Q}{\alpha_gN\ln 2}+1}^2-1}\sin^2\gamma+1} \times
\left\{ \begin{array}{cl}
        1, & \mathrm{j}=\mathrm{q}\\
        2, & \mathrm{j}=+\\
        4 Q^2\cos^2\gamma, & \mathrm{j}=\mathrm{E}\\
        \frac{1}{2} \frac{Q^2}{y^2} \sin^2\gamma, & \mathrm{j}=\mathrm{tr}\\
         \end{array} \right. ;
\label{B6}
\end{equation}
  \item $y\ll Q$ :
\begin{equation}
\fkt{\mathcal{I}_{3,2}^{\mathrm{j}}}{y,Q,\gamma}\simeq 
\frac{v_F^2}{T}\frac{1}{N}\frac{2\pi}{\ln 
2}\sqrt{y}\frac{Q^{-3/2}\cos\frac{\gamma}{2}}
{\displaystyle \frac{4y}{Q}\fkt{}{\frac{Q}{\alpha_gN\ln 
2}+1}^2\cos^2\frac{\gamma}{2}+1}\times
\left\{ \begin{array}{cl}
        1, & \mathrm{j}=\mathrm{q}\\ 
        2\sin^2\frac{\gamma}{2}, & \mathrm{j}=+\\
        4 Q^2, & \mathrm{j}=\mathrm{E}\\
        \frac{1}{2} \sin^2\gamma, & \mathrm{j}=\mathrm{tr}\\
         \end{array} \right. .
\label{B7}
\end{equation}
\end{itemize}
\subsubsection{Region 4}
This region is only important for the energy relaxation rate 
($\mathrm{j}=\mathrm{E}$). The corresponding contribution is governed by 
$Q\ll y$,
so that only the energy range $y\gg 1$ is of interest, where
\begin{equation}
\fkt{\mathcal{I}_{4,1}^{\mathrm{E}}}{y,Q,\gamma}\simeq
  \frac{v_F^2}{T}\frac{\alpha_g^2\pi^24N}{\sqrt{2\pi 
Q}}\frac{\sin\gamma\cos^2\gamma}{\displaystyle\fkt{}{\sin\gamma+\frac{\alpha_g\pi 
N}{8}}^2}
\end{equation}

\subsection{Results for the rates}
Finally we estimate the rates in the Regimes I, II, III, and IV as 
defined in Fig.~\ref{FigScales}, using the simplified integrands of the 
previous part.
\subsubsection{Region 1}
The contributions of Region 1 to the quantum and transport scattering 
rates are only relevant for energies in
Regime $\mathrm{II}$ for $y\ll Q$, where we have:
\begin{equation}
  \fkt{\tau^{-1}_{\mathrm{j}}}{y}=
\frac{2 
T^2}{v_F^2}\pint{Q}{\fkt{\mathrm{min}}{y,1}}{1}Q\pint{\gamma}{0}{2\pi}\fkt{\mathcal{I}_{1,2}^{\mathrm{j}}}{y,Q,\gamma}.
\end{equation}
Performing first the angular integration over $\gamma$,
we obtain the quantum scattering rate:
\begin{equation}
  \fkt{\tau^{-1}_\mathrm{q}}{y}\simeq 
4\alpha_gT\nint{a}{0}{1}
\frac{1}{\sqrt{1-a^2}}\simeq 2\pi\alpha_gT,
\end{equation}
the chiral scattering rate:
\begin{equation}
  \fkt{\tau^{-1}_+}{y}\simeq 
4\alpha_gT\nint{a}{0}{1}\, 
\sqrt{1-a^2}\simeq \pi\alpha_gT,
\end{equation}
and the transport scattering rate:
\begin{equation}
  \fkt{\tau^{-1}_\mathrm{tr}}{y}\simeq 
8\alpha_gT\nint{a}{0}{1}\, 
a^2\sqrt{1-a^2}\simeq \frac{\pi}{2}\alpha_gT. 
\end{equation}
Here the integrals over the variable $a=\alpha_g N \ln 2/\sqrt{4yQ}$ correspond to 
the $Q$ integration.

\begin{table}[H]
\begin{center}
  \begin{tabular}{|c|c|c|c|c|c|}
\hline
   \hspace*{0.5cm} Regime \hspace*{0.5cm} & 
\hspace*{0.8cm}$\fkt{\tau^{-1}_+}{y}$\hspace*{0.8cm} &\hspace*{1cm} Q 
\hspace*{1cm}&\hspace*{1cm} $\Omega$\hspace*{1cm} & 
\hspace*{1cm}$\gamma$ \hspace*{1cm}&\hspace*{1cm} $\theta$\hspace*{1cm} \\
\hline
   $\mathrm{II}$&$\simeq 2\pi\alpha_gT$ & $\alpha_g^2 N^2/y$& 
$ Q$ & $0<\gamma<\pi$ & $ 0<\theta<\pi $ \\
\hline
  \end{tabular}
\end{center}
\caption{Contribution of region 1 to the chiral quantum scattering rate $\tau_+^{-1}$ in Regime II 
and the characteristic values of
$Q$,$\Omega$, $\gamma$, and $\theta$ dominating this contribution.}
\end{table}

\begin{table}[H]
\begin{center}
  \begin{tabular}{|c|c|c|c|c|c|}
\hline
   \hspace*{0.5cm} Regime \hspace*{0.5cm} & 
\hspace*{0.8cm}$\fkt{\tau^{-1}_\mathrm{tr}}{y}$\hspace*{0.8cm} 
&\hspace*{1cm} Q \hspace*{1cm}&\hspace*{1cm} $\Omega$\hspace*{1cm} & 
\hspace*{1cm}$\gamma$ \hspace*{1cm}&\hspace*{1cm} $\theta$\hspace*{1cm} \\
\hline
   $\mathrm{II}$&$\simeq \pi\,\alpha_gT/2$ & $\alpha_g^2 N^2/y$ 
  & $Q$ & $ 0<\gamma<\pi$ & $ 0<\theta<\pi$ \\
\hline
  \end{tabular}
\end{center}
\caption{
Contribution of region 1 to the transport scattering rate $\tau_{\mathrm{tr}}^{-1}$ in Regime II 
and the characteristic values of $Q$,$\Omega$, $\gamma$, and $\theta$ dominating this contribution.}
\end{table}

In all the regimes, the contribution of region 1 to the energy 
relaxation rate is parametrically the same as that of other
regions. Since the numerical prefactor is not accessible within our 
calculation (the integrals are dominated by $qv_F\sim T$),
we have chosen to present only the calculation of the contribution of 
region 3.

\subsubsection{Region 2}
There is no important contribution from region 2 (see Sec.~\ref{Region2} 
above).
\subsubsection{Region 3}
This is the most important region for all the rates in most of the Regimes.
The corresponding integrals are expressed through the kernels
$ \mathcal{I}_{3,i}^{\mathrm{j}}$ introduced in 
Sec.~\ref{SectSimplIntegrands} as follows:
\begin{equation}
  \fkt{\tau^{-1}_{\mathrm{j}}}{y}
=\frac{2 
T^2}{v_F^2}\pint{Q}{0}{\fkt{\mathrm{min}}{y,1}}Q\pint{\gamma}{0}{2\pi}
\fkt{\mathcal{I}_{3,1}^{\mathrm{j}}}{y,Q,\gamma}+\frac{2 
T^2}{v_F^2}\pint{Q}{\fkt{\mathrm{min}}{y,1}}{1}Q
\pint{\gamma}{0}{2\pi}\fkt{\mathcal{I}_{3,2}^{\mathrm{j}}}{y,Q,\gamma}
\end{equation}
Using the  short-hand notations 
$$x'=\frac{1}{\alpha_gN\ln 2}, \quad  y'=\frac{y}{\alpha_gN\ln 2},$$
we obtain the following result for the contribution of region 3 to the quantum scattering rate:
\begin{equation}
                \fkt{\tau^{-1}_\mathrm{q}}{y}
\simeq 
\frac{\alpha_g}{\pi}T\fktc{}{\nint{x}{0}{\hspace*{-0.1cm}\fkt{\mathrm{min}}{x',y'}\hspace*{-0.3cm}}
\frac{\fktb{\mathrm{arsinh}}{\sqrt{x(x+2)}}}{\sqrt{x(x+2)}(x+1)}+\nint{x}{\hspace*{-0.3cm}\fkt{\mathrm{min}}{x',y'}
\hspace*{-0.3cm}}{x'}\frac{\fktb{\mathrm{arsinh}}{\sqrt{4y'}\fkt{}{\sqrt{x}+\frac{1}{\sqrt{x}}}}}{\fkt{}{1+x} 
\sqrt{1+4y'\fkt{}{\sqrt{x}+\frac{1}{\sqrt{x}}}^2}}}.
\end{equation}

\begin{table}[H]
\begin{center}
  \begin{tabular}{|c|c|c|c|c|c|}
\hline
   \hspace*{0.5cm} Regime \hspace*{0.5cm} & 
\hspace*{0.8cm}$\fkt{\tau^{-1}_\mathrm{q}}{y}$\hspace*{0.8cm} 
&\hspace*{1cm} Q \hspace*{1cm}&\hspace*{1cm} $\Omega$\hspace*{1cm} & 
\hspace*{1cm}$\gamma$ \hspace*{1cm}&\hspace*{1cm} $\theta$\hspace*{1cm} \\
\hline
   $\mathrm{I}$&$\sim \frac{\sqrt{y}}{N} T$ & $ 1$  & 
$ -Q$ & $ 0<\gamma<\pi$ & $ 0<\theta<\pi$ \\[0.2cm]
\hline
   $\mathrm{II}$&$\simeq \pi\alpha_g T$ & $ \alpha_g^2 N^2/y$  & 
$ -Q$ & $ 0<\gamma<\pi$ & $ 0<\theta<\pi$ \\[0.2cm]
\hline
   $\mathrm{III}$&$\simeq \frac{4\fkt{\beta}{2}}{\pi}\alpha_g T$ & 
$\alpha_g N T$  & $ Q \cos\gamma $ & $0<\gamma<\pi$ & $ 
\theta\alt \frac{\alpha_g N}{y}\ll 1$ \\[0.2cm]
\hline
   $\mathrm{IV}$&$\simeq \frac{4\fkt{\beta}{2}}{\pi}\alpha_g T$ & 
$\alpha_g N T$  & $ Q \cos\gamma $ & $0<\gamma<\pi$ & $ 
\theta\alt \frac{\alpha_g N}{y}\ll 1$ \\[0.2cm]
\hline
  \end{tabular}
\end{center}
\caption{Contribution of region 3 to the total quantum scattering rate $\tau_q^{-1}$ in all energy domains
and the corresponding characteristic values of
the transferred momentum $Q$, transferred frequency $\Omega$, the angle $\gamma$
between momenta $\vec{p}$ and $\vec{q}$, and the scattering angle $\theta$.}
\end{table}
For the contribution of region 3 to the chiral scattering rate we get:
\begin{multline}
                \fkt{\tau^{-1}_+}{y}
\simeq 
\frac{\alpha_g}{\pi}T\fktc{}{\nint{x}{0}{\hspace*{-0.1cm}\fkt{\mathrm{min}}{x',y'}\hspace*{-0.3cm}}
\frac{\fktb{\mathrm{arsinh}}{\sqrt{x(x+2)}}}{\sqrt{x(x+2)}(x+1)}\right. 
\\ \left. +\nint{x}{\hspace*{-0.3cm}\fkt{\mathrm{min}}{x',y'}
\hspace*{-0.3cm}}{x'}\frac{\sqrt{1+4y'\fkt{}{\sqrt{x}+\frac{1}{\sqrt{x}}}^2}\fktb{\mathrm{arsinh}}{\sqrt{4y'}\fkt{}{\sqrt{x}+\frac{1}{\sqrt{x}}}}-\sqrt{4y'}\fkt{}{\sqrt{x}+\frac{1}{\sqrt{x}}}}{\fkt{}{1+x}^2 
\frac{4y'}{x} }}
\end{multline}

\begin{table}[H]
\begin{center}
  \begin{tabular}{|c|c|c|c|c|c|}
\hline
   \hspace*{0.5cm} Regime \hspace*{0.5cm} & 
\hspace*{0.8cm}$\fkt{\tau^{-1}_+}{y}$\hspace*{0.8cm} 
&\hspace*{1cm} Q \hspace*{1cm}&\hspace*{1cm} $\Omega$\hspace*{1cm} & 
\hspace*{1cm}$\gamma$ \hspace*{1cm}&\hspace*{1cm} $\theta$\hspace*{1cm} \\
\hline
   $\mathrm{I}$&$\sim \frac{\sqrt{y}}{ N} T$ & $1$ 
& $-Q$ & $ 0<\gamma<\pi$ & $ 0<\theta<\pi$ \\[0.2cm]
\hline
   $\mathrm{II}$&$\simeq \pi \alpha_g T$ & $ \alpha_g^2 
N^2/y$  & $ -Q$ & $ 0<\gamma<\pi$ & $ 0<\theta<\pi$ \\[0.2cm]
\hline
   $\mathrm{III}$&$\simeq \frac{8\fkt{\beta}{2}}{\pi}\alpha_g T$ & 
$\alpha_g N T$  & $ Q \cos\gamma $ & $0<\gamma<\pi$ & $ 
\theta\alt \frac{\alpha_g N}{y}\ll 1$ \\[0.2cm]
\hline
   $\mathrm{IV}$&$\simeq \frac{8\fkt{\beta}{2}}{\pi}\alpha_g T$ & 
$ \alpha_g N T$  & $ Q \cos\gamma $ & $0<\gamma<\pi$ & $ 
\theta\alt \frac{\alpha_g N}{y}\ll 1$ \\[0.2cm]
\hline
  \end{tabular}
\end{center}
\caption{Contribution of region 3 to the chiral quantum scattering rate $\tau_+^{-1}$ in all energy domains
and the corresponding characteristic values of
$Q$, $\Omega$, $\gamma$, and $\theta$.}
\end{table}

The contribution of region 3 to the energy relaxation rate reads:
\begin{equation}
                \fkt{\tau^{-1}_\mathrm{E}}{y}
\simeq\frac{\alpha_g}{\pi (x')^2} T
\fktc{}{\nint{x}{0}{\hspace*{-0.1cm}\fkt{\mathrm{min}}{x',y'}\hspace*{-0.3cm}}
\frac{x}{x+2}\fktb{}{\frac{\fkt{\mathrm{arsinh}}{\sqrt{x(x+2)}}}{(x+1)^{-1}\sqrt{x(x+2)}}-1} 
+\nint{x}{\hspace*{-0.3cm}\fkt{\mathrm{min}}{x',y'}
\hspace*{-0.3cm}}{x'}\frac{x^2\fktb{\mathrm{arsinh}}{\sqrt{4y'}
\fkt{}{\sqrt{x}+\frac{1}{\sqrt{x}}}}}{\fkt{}{1+x} 
\sqrt{1+4y'\fkt{}{\sqrt{x}+\frac{1}{\sqrt{x}}}^2}}}. 
\end{equation}

\begin{table}[H]
\begin{center}
  \begin{tabular}{|c|c|c|c|c|c|}
\hline
   \hspace*{0.5cm} Regime \hspace*{0.5cm} & 
\hspace*{0.8cm}$\fkt{\tau^{-1}_\mathrm{E}}{y}$\hspace*{0.8cm} 
&\hspace*{1cm} Q \hspace*{1cm}&\hspace*{1cm} $\Omega$\hspace*{1cm} & 
\hspace*{1cm}$\gamma$ \hspace*{1cm}&\hspace*{1cm} $\theta$\hspace*{1cm} \\
\hline
   $\mathrm{I}$&$\sim \frac{\sqrt{y}}{ N} T$ & $ 1$ 
  & $ -Q$ & $ 0<\gamma<\pi$ & $ 0<\theta<\pi$ \\[0.2cm]
\hline
   $\mathrm{II}$&$\sim \alpha_g^2 T \frac{N}{\sqrt{y}}\fkt{\log}{\frac{y}{\alpha_g^2N^2}} 
$ & $ 1$  & $ -Q$ & $  
\pi-\frac{\alpha_gN}{\sqrt{y}}$ & $  \frac{\alpha_gN}{\sqrt{y}}\ll 1$ 
\\[0.2cm]
\hline
   $\mathrm{III}$& $\sim \alpha_g^2 T \frac{N}{\sqrt{y}}\fkt{\log}{\frac{y}{\alpha_g^2N^2}} 
$ & $ 1$  & $ -Q $ & $  
\pi-\frac{\alpha_gN}{\sqrt{y}}$ & $  \frac{\alpha_gN}{\sqrt{y}}\ll 1$ 
\\[0.2cm]
\hline
   $\mathrm{IV}$&$\sim  \alpha_g^2 N T \fkt{\log}{\frac{1}{\alpha_g^2N^2}}$ & $ 1$  & $ -Q  $ & $  
\pi-\frac{\alpha_gN}{\sqrt{y}}$ & $  \theta\alt \frac{\alpha_g N}{y}\ll 1$ 
\\[0.2cm]
\hline
  \end{tabular}
\end{center}
\caption{Contribution of region 3 to the energy relaxation rate $\tau_\mathrm{E}^{-1}$ obtained in all energy domains
within the energy-diffusion approximation and the corresponding characteristic values of
$Q$, $\Omega$, $\gamma$, and $\theta$. Since in all energy domains the result is dominated by $Q\sim 1$, the numerical prefactors
can not be obtained from the asymptotics at $Q\ll 1$ and $Q\gg 1$.}
\end{table}

Finally, region 3 yields the following result for $\tau^{-1}_\mathrm{tr}$:
\begin{multline}
\fkt{\tau^{-1}_\mathrm{tr}}{y}
\simeq\frac{\alpha_g}{2\pi 
y'^2}T\hspace*{-0.3cm}\nint{x}{0}{\hspace*{-0.3cm}\fkt{\mathrm{min}}{x',y'}
\hspace*{-0.3cm}}\frac{x}{x+2}\fktb{}{1-\frac{\fkt{\mathrm{arsinh}}{\sqrt{x(x+2)}}}{(x+1)\sqrt{x(x+2)}}} 
\\ +\frac{4}{\pi}\alpha_g 
T\hspace*{-0.2cm}\nint{x}{\hspace*{-0.35cm}\fkt{\mathrm{min}}{x',y'}\hspace*{-0.35cm}}{x'} 
\sqrt{\frac{y'}{x}}\fktb{}{\frac{z^2+3}{3 
z^4}-\frac{\mathrm{arsinh}z}{z^5\fkt{}{1+z^2}^{-1/2}}}_{z=\sqrt{4y'}\fkt{}{\sqrt{x}+\frac{1}{\sqrt{x}}}}.
\end{multline}

\begin{table}[H]
\begin{center}
  \begin{tabular}{|c|c|c|c|c|c|}
\hline
   \hspace*{0.5cm} Regime \hspace*{0.5cm} & 
\hspace*{0.8cm}$\fkt{\tau^{-1}_\mathrm{tr}}{y}$\hspace*{0.8cm} 
&\hspace*{1cm} Q \hspace*{1cm}&\hspace*{1cm} $\Omega$\hspace*{1cm} & 
\hspace*{1cm}$\gamma$ \hspace*{1cm}&\hspace*{1cm} $\theta$\hspace*{1cm} \\
\hline
   $\mathrm{I}$&$\sim \frac{\sqrt{y}}{N} T$ & $ 1$ 
  & $ -Q$ & $ 0<\gamma<\pi$ & $ 0<\theta<\pi$ \\[0.2cm]
\hline
   $\mathrm{II}$&$\simeq \frac{\pi}{4}\alpha_g T$ & $ 
\frac{\alpha_g^2N^2}{y}$  & $ -Q$ & $  
\pi-\frac{\alpha_gN}{\sqrt{y}}$ & $  \frac{\alpha_gN}{\sqrt{y}}\ll 1$ 
\\[0.2cm]
\hline
   $\mathrm{III}$&$\sim \frac{N}{y}\alpha_g^2 T$ & 
$ y$  & $ y(1-2|\sin\frac{\gamma}{2}|) $ & $  
\frac{\alpha_gN}{y}$ \& $  \pi-\frac{\alpha_gN}{y}$  & $  
\theta\alt \frac{\alpha_g N}{y}\ll 1$\\[0.2cm]
\hline
   $\mathrm{IV}$&$\sim \frac{N}{y^2}\alpha_g^2 T$ & $ 
1$  & $ Q \cos\gamma  $ & $  \frac{\alpha_gN}{y}$  & $\theta\alt \frac{\alpha_g N}{y}\ll 1$ \\[0.2cm]
\hline
  \end{tabular}
\end{center}
\caption{Contribution of region 3 to the transport scattering rate $\tau_\mathrm{tr}^{-1}$  in all energy domains
 and the corresponding characteristic values of
$Q$, $\Omega$, $\gamma$, and $\theta$. Since in energy domains I and IV the result is dominated by $Q\sim 1$, the numerical prefactors
in these domains are beyond the accuracy of our approximations. Furthermore, since in Regime III
the integral is dominated by $Q\sim y$, the splitting of the integrand according Eqs.~\eqref{B6}
and \eqref{B7} does not reproduce the correct prefactor.}
\end{table}

\subsubsection{Region 4}
This region is only relevant for the energy relaxation rate in Regime 
$\mathrm{IV}$, where the main
contribution comes from $Q\ll y$, yielding
\begin{equation}
  \fkt{\tau^{-1}_{\mathrm{E}}}{y}
=\frac{8 T^2}{v_F^2}\pint{Q}{1}{\fkt{\mathrm{max}}{y,1}}Q
\pint{\gamma}{0}{\frac{\pi}{2}-\frac{1}{Q}}\fkt{\mathcal{I}_{4,1}^{\mathrm{E},\text{ 
$\gamma \ll \frac{\pi}{2}-\frac{1}{Q}$}}}{y,Q,\gamma}\approx 
\frac{12N}{\sqrt{2\pi}}\, 
\alpha_g^2Ty^{3/2}\fkt{\log}{\frac{16}{N\pi\alpha_g\sqrt{y}}}.
\end{equation}
Since the result is determined by $Q\sim \Omega\sim y\gg 1$, the energy-diffusion model is, in fact, not applicable in this 
Regime.

\section{Scattering Rates from the Generalized Golden Rule}
\label{GR}

In this Appendix we analyze the golden rule approach to calculating quantum and transport scattering rates. The quantum scattering rate for electrons was introduced in Eq.\ (\ref{tauPMdef}). Using the imaginary part of the self energy Eq.\ (24), we obtain
\begin{multline}   \label{EqMatrixelement}
\fkt{\tau_{+}^{-1}}{\epsilon} \propto
\int d\vec{q}\int d\omega\;
\ABS{\fkt{D_{\mathrm{RPA}}}{q,\omega}}^2
[1-n_F(\epsilon-\omega)]
\int d\vec{p}
\mathop{\mathrm{Tr}} \left[ \fkt{\mathcal{P}_{+}}{\vec{p}}\fkt{\delta}{\epsilon-v_F p} \mathcal{A}_0(\epsilon-\omega,\vec{p}-\vec{q}) \right] \\ 
\times  
\int dE\;
\fktb{}{\fkt{n_F}{E-\omega/2}-\fkt{n_F}{E+\omega/2}}
\int d\vec{k} \mathop{\mathrm{Tr}} \left[
\mathcal{A}_0(E-\omega/2,\vec{k}-\vec{q}/2) \mathcal{A}_0(E+\omega/2,\vec{k}+\vec{q}/2)
\right].
\end{multline}
The spectral weights $\mathcal{A}_0$ are defined in Eq.\ (\ref{GlSpektralesgewicht}). Equation (\ref{EqMatrixelement}) is completely equivalent to Eq.\ (46) of the main text.

\begin{figure}
\begin{center}
  \includegraphics[width=5cm]{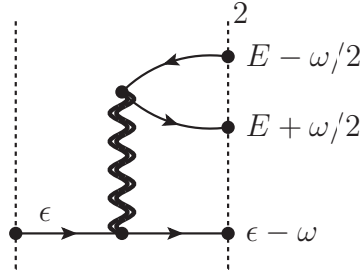}
\caption{Diagram for the ``elementary'' inelastic scattering amplitude.}
\label{Figdecayprocess}
\end{center}\end{figure}

We can regard the quantum scattering rate as the probability (per unit time) of the electron decay with emitting an electron-hole pair. The amplitude of this decay process is given by the "half" of the self-energy diagram shown in Fig.\ \ref{Figdecayprocess}. The incoming and outgoing particles in this diagram are taken at the mass shell. Equation (\ref{EqMatrixelement}) has the form of the Fermi golden rule with the amplitude determined by the RPA-screened interaction.

We can now apply the golden rule to calculate the transport scattering rate. This amounts to including an extra transport factor
$[\vec{v}_i\cdot (\vec{v}_i-\vec{v}_f)]/v_F^2$ accounting for the change of the current due to scattering, in the integrand of Eq.\ (\ref{EqMatrixelement}). Here $\vec{v}_i$ and $\vec{v}_f$ are total velocities of incoming and outgoing particles. 
For the linear electronic dispersion the velocities of individual quasiparticles are determined by the relation 
\begin{equation} \label{velocity-momentum}
\vec{v}=\frac{\partial\epsilon}{\partial\vec{p}}=\frac{v_F^2\vec{p}}{\epsilon}
\end{equation}
This results in the following transport factor:
\begin{equation} \label{transport-factor}
\frac{\vec{v}_i\cdot\fkt{}{\vec{v}_i-\vec{v}_f}}{v_F^2}
 = v_F^2 \frac{\vec{p}}{\epsilon}\fkt{}{
     \frac{\vec{p}}{\epsilon}
     -\frac{\vec{p}-\vec{q}}{\epsilon-\omega}
     -\frac{\vec{k}+\vec{q}/2}{E+\omega/2}
     +\frac{\vec{k}-\vec{q}/2}{E-\omega/2}
   },
\end{equation}
Note that in the conventional case of massive particles with the quadratic electronic dispersion, $\vec{v}=\vec{p}/m$ and hence
$\vec{v}_i-\vec{v}_f\propto \vec{p} - (\vec{p}-\vec{q})-(\vec{k}+\vec{q}/2)+(\vec{k}-\vec{q}/2)=0$, which reflects the
fact that because of the total momentum conservation there is no current relaxation in conventional metals due to the electron-electron interaction.

We now apply the particle-hole symmetry of the graphene spectrum in order to simplify the expression for transport scattering rate. Let us reverse the integration variables in the second line of Eq.\ (\ref{EqMatrixelement}): $\fkt{}{E,\vec{k}} \mapsto \fkt{}{-E,-\vec{k}}$. Using the symmetry $\mathcal{A}_0(\epsilon,\vec{p}) = \mathcal{A}_0(-\epsilon,-\vec{p})$, we see that the trace in the integrand is not changed. The difference of two equilibrium distribution functions in the second line of Eq.\ (\ref{EqMatrixelement}) is also independent of the sign of $E$. 
Thus the contribution of particle-hole pair into the current relaxation vanishes, which corresponds to the absence 
of the Coulomb-drag contribution to the conductivity~\cite{GM} in undoped graphene (Fig. \ref{FigCondDiagssecond}f).
This allows us to keep only the part of the transport factor Eq.\ (\ref{transport-factor}) which is even under reversing $E$ and $\vec{k}$,
\begin{equation}
 v_F^2 \frac{\vec{p}}{\epsilon}\fkt{}{
   \frac{\vec{p}}{\epsilon} - \frac{\vec{p}-\vec{q}}{\epsilon-\omega}
 } = 1 - \mathop{\mathrm{sign}}(\epsilon-\omega) \cos\theta,
\end{equation}
where $\theta$ is the scattering angle. Inserting this reduced transport factor into the first line of Eq.\ (\ref{EqMatrixelement}) and calculating the trace of projection operators, we obtain
\begin{multline}
[1 - \mathop{\mathrm{sign}}(\epsilon-\omega) \cos\theta]
\mathop{\mathrm{Tr}} \left[ \mathcal{P}_{+}(\vec{p}) \mathcal{A}_0(\epsilon-\omega, \vec{p}-\vec{q}) \right] \\
= \frac{1}{2} \big[ 1 - \mathop{\mathrm{sign}}(\epsilon-\omega) \cos\theta \big]
\big[
(1+\cos\theta) \delta(\epsilon-\omega-v_F|\vec{p}-\vec{q}|)
+(1-\cos\theta) \delta(\epsilon-\omega+v_F|\vec{p}-\vec{q}|)
\big] \\ = \frac{1}{2}
(1-\cos^2\theta) \big[ \delta(\epsilon-\omega-v_F |\vec{p}-\vec{q}|) + \delta(\epsilon-\omega+v_F|\vec{p}-\vec{q}|) \big].
\end{multline}
This way the transport factor $1 - \cos^2\theta$ appears for Dirac fermions.

The golden rule calculation of the transport scattering rate reproduces the result obtained from the Drude conductivity Eq.\ (\ref{EqCondCorrwithReg}). Note that an additional broadening $\delta$ was not used in this calculation. This shows that our result for $\tau_{\mathrm{tr}}^{-1}$ can be actually applied in the limit $\tau_{\mathrm{tr}}^{-1} \gg \delta$ as we argued in Sec. \ref{SubSectCollisionLimitedC}.

\end{appendix}

\twocolumngrid

\end{document}